\documentclass[a4paper,12pt]{article}
\usepackage[utf8]{inputenc}
\usepackage[english]{babel}
\usepackage{url}
\usepackage[T1]{fontenc}
\usepackage{amsmath}
\usepackage{graphicx} 
\usepackage{apacite}
\usepackage[round,authoryear]{natbib}
\usepackage{authblk}
\usepackage{csquotes}
\usepackage{rotating}
\usepackage{setspace}
\usepackage{fancyhdr}
\usepackage[a4paper,top=2cm,bottom=2cm,left=2cm,right=2cm]{geometry} 
\providecommand{\keywords}[1]{\textbf{Keywords: }#1}
\providecommand{\jel}[1]{\textbf{JEL Classification: }#1}

\author[1,2]{Jacek Białek}
\affil[1]{University of~Łódź, Department of~Statistical Methods, Łódź, Poland, jacek.bialek@uni.lodz.pl}
\affil[2]{Statistics Poland, Department of~Trade and Services, Poland, J.Bialek@stat.gov.pl}

\author[3,4]{Maciej Beręsewicz}
\affil[3]{Poznań University of~Economics and Business, Department of~Statistics, Poland, maciej.beresewicz@ue.poznan.pl}
\affil[4]{Statistical Office in~Poznań, Centre for~Small Area Estimation, Poland, m.beresewicz@stat.gov.pl}

\date{}
\title{Scanner data in inflation measurement: from raw data to price indices}
\newcommand{\bbeta}{\mathbf{\beta}}

\newcommand{\bY}{\mathbf{Y}}

\newcommand\blfootnote[1]{%
  \begingroup
  \renewcommand\thefootnote{}\footnote{#1}%
  \addtocounter{footnote}{-1}%
  \endgroup
}

\begin{document}

\maketitle 
\thispagestyle{empty}

\begin{abstract} 
\footnotesize
Scanner data offer new opportunities for CPI or HICP calculation. They can be obtained from a~wide variety of~retailers (supermarkets, home electronics, Internet shops, etc.) and provide information at the level of~the barcode. One of~advantages of~using scanner data is the fact that they contain complete transaction information, i.e. prices and quantities for every sold item. To use scanner data, it must be carefully processed. After clearing data and unifying product names, products should be carefully classified (e.g. into COICOP 5 or below), matched, filtered and aggregated. These procedures often require creating new IT or writing custom scripts (R, Python, Mathematica, SAS, others). One of~new challenges connected with scanner data is the appropriate choice of~the index formula. In this article we present a~proposal for the implementation of~individual stages of~handling scanner data. We also point out potential problems during scanner data processing and their solutions. Finally, we compare  a~large number of~price index methods based on real scanner datasets and we verify their sensitivity on adopted data filtering and aggregating methods.
\setstretch{1.5}
\end{abstract}

\noindent \keywords{Scanner data, product classification, product matching, Consumer Price Index, multilateral indices}

\noindent  \jel{C43, E31}

\blfootnote{\textbf{Acknowledgements}: This publication if financed by the National Science Centre in Poland (grant no. 2017/25/B/HS4/00387).  The views expressed are those of~the authors and not necessarily those of Statistics Poland.}







\clearpage
\setstretch{1.5}

\section{Introduction}

The term “scanner data” refers to transaction data that specify turnover and numbers of~items sold by GTIN (barcode, formerly known as the EAN code) or other barcodes (like SKU, PLU or UPC). Scanner data have numerous advantages compared to traditional survey data: for one thing, such data sets are much bigger than traditional ones and contain complete transaction records, i.e. information about prices and quantities. In other words, scanner data contain expenditure information at item level (i.e. at barcode or GTIN level). Consequently, expenditure shares of~items can be used as weights for calculating price indices at the lowest (elementary) level of~data aggregation. 

In 2002 scanner data from two supermarkets were introduced in the Dutch CPI and, in January 2010, the number of~supermarkets providing scanner data rose to six. As a~result, the Dutch CPI was re-designed \citep{de2006re, grient2010, de2011eliminating}. In 2017, scanner data of~ten supermarkets chains were used and at present surveys are no longer carried out for supermarkets; instead, scanner data from other retailers (for instance, from do-it-yourself stores or from travel agencies) are used to calculate the Dutch CPI \citep{chessa2015towards}. Until  2015, only four EU countries took advantage of~scanner data (the Netherlands, Norway, Sweden, and Switzerland). Since then, the number of~countries that make use of~this data source for calculating their CPI has been growing: in April 2016, three more countries followed suit (Belgium, Sweden and Switzerland) and at present, some national statistical institutes (NSIs) are also considering this possibility. For instance, in 2010 the French National Statistical Institute (INSEE) launched a~pilot project to investigate the suitability of~scanner data for CPI calculation; in 2011 Statistics Portugal was awarded a~Eurostat grant to undertake initial research on the exploitation of~scanner data; in Luxembourg, collaboration was established with several retailers, who agreed to send their monthly data to the country's NSI (STATEC), enabing the use of~scanner data in regular production from January 2018. In January 2018, Statistics Poland launched a~project called “INSTATCENY”, whose main aim is to create a~new methodology of~CPI measurement based on data from different (traditional and non-traditional) sources, including scanner data and web-scraped data. In 2017, Eurostat published Practical Guide for Processing Supermarket Scanner Data, which is available at \url{https://ec.europa.eu/eurostat/web/hicp/overview}). In the Foreword to the guide its authors note: “This guide describes the situation in 2017. It will need to be updated as the use of~scanner data develops and broadens”. In fact, the methodology for CPI (or HICP) construction using scanner data has strongly evolved over the last few years (see for instance: \citet{krsinich2014fews, ivancic2011scanner, Griffioen2016dutch, de2016overview, chessa2016comparing, chessa2017comparisons, diewert2018substitution, Auer2019, bialek2019, Webster2019,mehrhoff2019,  Zhang2019}. One of~the new challenges connected with scanner data is the choice of~the index formula which should be able to reduce the chain drift bias and the substitution bias. 

In this article we present a~proposal for implementing individual stages of~handling scanner data. We also point out potential problems during scanner data processing and possible ways of~solving them. The article is organised as follows: Section \ref{scanner-data} presents the main challenges faced when using scanner data. Section \ref{section-best-price-index} describes one bilateral and multilateral index method which can be applied to scanner data and contains a~discussion of~the updating and weighting problem connected with multilateral methods. Section \ref{prepare-data} describes how to prepare scanner data sets for further analysis and how to deal with product classification by using machine learning methods; this section also presents theoretical assumptions and our results of~product matching. Section \ref{sec-filtering} considers the usefulness of~data filters before index calculations. Section \ref{sec-comparison} compares the selected price index methods taking into account the impact of~aggregating across outlets and subgroups of~products. Section \ref{sec-conclus} contains the main conclusions.

\section{Scanner data: challenges at the stage of~data processing}\label{scanner-data}

One of~the main advantages of~using scanner data is the possibility of~aggregation at lower levels, since information about prices and quantities (and consequently about weights) is available. Scanner data sets are huge and can provide some additional information about products (including such attributes as size, colour, package quantity, etc.). These attributes are useful for aggregating items into homogeneous groups. In addition, scanner data are much cheaper to obtain than survey data traditionally used for CPI calculation. As pointed out in Eurostat’s  Practical Guide for Processing Supermarket Scanner Data from 2017: “In the traditional price collection, price collectors have to trust intuition and common sense and it may happen that prices are collected as long as the item is available even though it is no longer representative. In scanner data the representativeness is guaranteed” (\citet{eurostat2017}:9). 

Generally speaking, there are many challenges involved in scanner data processing. The first challenge concerns item codes. As mentioned above, the Global Trade Item Number (GTIN) (formerly known as EAN) is currently used for coding scanner data. Nevertheless, the following codes can also be used: price look-up (PLU) and stock-keeping units (SKUs). PLU codes are shorter than GTINs, and SKUs can be slightly more generic than GTINs. In practice, when NSIs use scanned data from different retailers with different code systems, there may be some problems with identifying products. Having made sure that all data represent accepted code systems (retailer codes and/or external bar codes), after data cleaning and unifying product names, products must be carefully classified (e.g. into COICOP 5 categories or even lower), matched and filtered. The list of~challenges does not end here: scanner data may contain data about transactions between the retailer and other businesses, which should be verified and detected (such transactions should be excluded from CPI calculations). Another challenge involves detecting items which were returned within a~given period after the purchase. Since a~typical supermarket uses 10000-25000 item codes, what is required is an IT system capable of~detecting such items automatically or almost automatically, taking into consideration seasonal goods, replacements, as well as item codes that disappear and appear in the sample.

Statistics Poland has been cooperating with 2 supermarket chains for almost 4 years but only one of~them provides scanner data on a~regular basis (currently negotiations are under way with 4 other retail chains specialising in different product categories and this phase is scheduled to end in April 2020). Initially, scanner data were analysed in terms of~usability, representativeness and quality. About a~year ago, Statistics Poland began experimenting with automatic product classification, product matching and inflation measurement based on scanner data. Although Statistics Poland does not have solutions for all the problems and challenges discussed above, some of~them have been solved. The following sections discuss our most important achievements in this area and recommended solutions to selected problems.

\section{Scanner data: challenges with the choice of~the best price index formula}\label{section-best-price-index}

There are many price index methods which can be used for scanner data given that expenditure data are available at elementary level \citep{chessa2017comparison}. Before discussing different approaches to choosing a~price index for scanner data, let us first introduce price indices currently used and taken into consideration by Statistics Poland. Let us denote sets of~homogeneous products belonging to the same product group in months $0$ and $t$ by $G_0$ and $G_t$ respectively, and let $G_{0,t}$ denote a~set of~matched products in both moments $0$ and $t$.  We also assume that scanner data are aggregated over the period of~a~given month.

\subsection{Bilateral index methods}

\subsubsection{Unweighted formulas}

The following recommendation of~the European Commission concerning the choice of~the elementary formula at the lowest level of~data aggregation can be found at \url{http://www.ilo.org/public/english/bureau/stat/download/cpi/corrections/annex1.pdf}: “For the HICPs the ratio of~geometric mean prices or the ratio of~arithmetic mean prices are the two formulae which should be used within elementary aggregates. The arithmetic mean of~price relatives may only be applied in exceptional cases and where it can be shown that it is comparable”. In other words, if expenditure information is not available, the European Commission recommends the \citet{jevons1865variation} price index (see also \citet{diewert2012consumer} or \citet{levell2015carli}), which can be written as follows:

\begin{equation}
P_{J}^{0, t}=\prod_{i \in G_{0, t}}
\left(\frac{p_{i}^{t}}{p_{i}^{0}}\right)^{\frac{1}{N_{0, t}}},
\end{equation}

\noindent where $p_i^{\tau}$ denotes the price of~the $i$-th product at time $\tau \in \{0,t\}$ and $N_{0,t}= \operatorname{card} G_{0,t}$. Theoretically, for a~homogeneous group of~products we could also use the Dutot ($P_D^{0,t}$) and the Carruthers-Sellwood-Ward-Dalen (CSWD) price indices \citep{carruthers1980recent}. The Dutot index (1738) can be expressed as follows (see also the CPI Manual, \citeyear{oecd2004consumer}):

\begin{equation}
P_{D}^{0, t}=\frac{\sum\limits_{i \in G_{0, t}} p_{i}^{t}}{\sum\limits_{i \in G_{0, t}} p_{i}^{0}},
\end{equation}

\noindent and the CSWD index is a~geometric mean of~the Carli index (1804) and the harmonic mean of~price relatives \citet{lippe2007}. In our study, we consider all these bilateral formulas together with their monthly chained versions.

\subsubsection{Weighted formulas}

Since scanner data contain information about expenditures, they can be used to calculate weighted bilateral indices. Superlative price indices, firstly proposed by \citet{diewert1976exact}, are the most frequently recommended index formulas for scanner data (as base formulas). Following \citet{chessa2017comparisons}, we consider the \citet{tornqvist1936} price index, which is given by

\begin{equation}
P_{T}^{0, t}=\prod\limits_{i \in G_{0, t}}
\left(\frac{p_{i}^{t}}{p_{i}^{0}}\right)^{\frac{s_{1}^{0}+s_{i}^{t}}{2}},
\label{tornqvist1936eq}
\end{equation}

\noindent where $s_i^0$ and $s_i^t$ denote the expenditure shares of~matched products in months $0$ and $t$.  Another commonly known superlative price index is the Fisher price index (\citeyear{fisher1922making}), which can be written as:

\begin{equation}
P_{F}^{0, t}=\sqrt{P_{La}^{0, t} \cdot P_{Pa}^{0, t}},
\end{equation}

\noindent where $q_i^0$ and $q_t^i$  denote quantities of~matched products in months $0$ and $t$, $P_{La}^{0, t}$ and $P_{Pa}^{0, t}$  denote the Laspeyres price index (\citeyear{laspeyres1871ix}) and the Paasche price index (\citeyear{hp1874ueber}) respectively. Some statisticians use the Sato-Vartia index \citet{sato1976ideal, Vartia1976} for scanner data. We include this index in our study on account of~its good axiomatic properties \citep{lippe2007,abe2019multilateral}. Following \citep{sato1976ideal}, the log-change price index has the following generic form:

\begin{equation}
\ln P_{S V}^{0, t}=\sum_{i=1}^{N} \phi_{i}^{0, t} \ln \left(\frac{p_{i}^{t}}{p_{i}^{s}}\right),
\end{equation}

\noindent where weights $\phi_{i}^{0, t}$ are based on logarithmic means of~expenditures shares $s^0_i$ and $s^t_i$, i.e. 

\begin{equation}
\phi_{i}^{0, t}=\frac{
\frac{s_{i}^{t}-s_{i}^{0}}{\ln \left(s_{i}^{t}\right)-\ln \left(s_{i}^{0}\right)}}{
\sum\limits_{k \in G_{0, t}} 
\frac{s_{k}^{t}-s_{k}^{0}}{\ln \left(s_{k}^{t}\right)-\ln \left(s_{k}^{0}\right)}}.
\end{equation}

\subsection{Multilateral index methods}

Multilateral index methods originate in comparisons of~price levels across countries or regions. These methods satisfy the condition of~transitivity, which is a~desirable property for spatial comparisons because the results are independent of~which country is selected as the base country (region). Commonly known methods include the GEKS method (see \citet{gini1931circular,elteto1964problem,geary1958note,khamis1972new}), the CCDI method \citep{caves1982multilateral} or the Time Product Dummy Methods \citep{de2018time}.

\subsubsection{The quality adjusted unit value index and the Geary-Khamis (GK) method}

The term “Quality adjusted unit value method” (QU method for short) was introduced by Chessa (see, for instance, \citet{chessa2015towards, chessa2016new}). The QU method is a~family of~unit value based index methods with the above-mentioned Geary-Khamis (GK) method as a~special case. According to the QU method, the price index  $P_{QU}^{0,t}$, which compares period $t$ with base period $0$, is defined as follows:

\begin{equation}
P_{Q U}^{0, t}=
\frac{
\sum\limits_{i \in G_{i}} p_{i}^{t} q_{i}^{t} / 
\sum\limits_{E C_{0}} p_{i}^{0} q_{i}^{0}}{
\sum\limits_{i \in G_{t}} v_{i} q_{i}^{t} / 
\sum\limits_{i \in G_{0}} v_{i} q_{i}^{0}},
\label{eq-gk}
\end{equation}

\noindent where the numerator in \eqref{eq-gk} is the measure  of~the turnover (expenditure) change between the two months and the denominator in \eqref{eq-gk} is a~weighted quantity index. Note that both the turnover index and the weighted quantity index are transitive, and thus the price index $P^{0,t}_{GU}$ is also transitive \citep{chessa2017comparison}. Note also that the quantity weights $v_i$ are the only unknown factors in formula \eqref{eq-gk} and these factors convert sold quantities $q_i^0$ and $q_{i}^t$ into “common units” $v_iq_i^0$ and $v_iq_i^t$. Prices of~products, $p_i^0$ and $p_i^t$ are converted into “quality adjusted prices” $p_i^0 / v_i$ and $p_i^t / v_i$. Different choices of~factors $v_i$ lead to different price index formulas. In the GK method, the weights $v_i$ are defined as follows: 

\begin{equation}
v_{i}=\sum_{z=0}^{T} \varphi_{i, G K}^{z} \frac{p_{i}^{z}}{P_{Q U}^{0, z}},
\end{equation}

\noindent  where

\begin{equation}
\varphi_{i, G K}^{z}=\frac{q_{i}^{z}}{\sum_{\tau=0}^{T} q_{i}^{\tau}},
\end{equation}

\noindent and $[0,T]$ is the entire time interval of~product observations (typically $T=12$, see \citet{diewert2018substitution}). Note that formulas (7), (8) and (9) lead to a~set of~equations which should be solved simultaneously. The above-mentioned solution can be found iteratively \citep{maddison1996generalized,chessa2016new} or as a~solution to an eigenvalue problem \citep{diewert1999axiomatic}. An~interesting alternative method for obtaining this solution can be also found in \citet{diewert2018substitution}.


\subsubsection{The GEKS method}

Let us consider a~time interval $[0,T]$ of~observations of~prices and quantities that will be used for constructing the GEKS index. The GEKS price index between months $0$ and $t$ is an unweighted geometric mean of~$T+1$ ratios of~bilateral price indices $P^{\tau, t}$ and $P^{\tau, 0}$, which are based on the same price index formula. The bilateral price index formula should satisfy the time reversal test, i.e. it should satisfy the condition $P^{a,b} \cdot P^{b,a} = 1$. Typically, the GEKS method uses the superlative Fisher price index, resulting in the following formula:

\begin{equation}
P_{G E K S}^{0, t}=\prod_{\tau=0}^{T}\left(\frac{P_{F}^{\tau, t}}{P_{F}^{\tau, 0}}\right)^{\frac{1}{T+1}}.
\end{equation}

\subsubsection{The CCDI method}

The GEKS method for making international index number comparisons comes from \citet{gini1931circular} but it was derived in a~different manner by \citet{elteto1964problem} and \citet{szulc1964indices}. \citet{feenstra2009consistent}, and also \citet{de2011eliminating} suggested that the T\"ornqvist price index formula (see \eqref{tornqvist1936eq}) could be used instead of~the Fisher price index in the Gini methodology. \citet{caves1982multilateral} used the GEKS idea with the T\"ornqvist index as a~base in the context of~making quantity comparisons across production units (the CCD method) and \citet{inklaar2016measuring} extended the CCD methodology to making price comparisons across production units. Consequently, in the article by \citet{diewert2018substitution}, the multilateral price comparison method involving the GEKS method based on the T\"ornqvist price index is called the CCDI method. The corresponding CCDI price index can be expressed as follows:

\begin{equation}
P_{C C D I}^{0, t}=\prod_{\tau=0}^{T}\left(\frac{P_{T}^{\tau, t}}{P_{T}^{\tau, 0}}\right)^{\frac{1}{T+1}}.
\end{equation}

\subsubsection{The TPD  method}

The Time Product Dummy method (TPD) uses a~panel regression approach to estimating price indices using all available data from a~given reference period. The econometric model for log prices observed during the time interval $[0,T]$ is as follows:

\begin{equation}
\ln p_{i}^{t}=\alpha+\sum_{t=1}^{T} \delta^{t} D_{i}^{t}+\sum_{i=1}^{N-1} \gamma_{i} D_{i}+\varepsilon_{i}^{t},
\end{equation}

\noindent where $D_i$ is the dummy variable that has value 1 if the $i$-th product is available in period $t$ and $0$ otherwise, $\delta^t$ are the time dummy parameters, $\gamma^t$ represents the item fixed effects and $\epsilon_i^t$ denotes the corresponding random error term. The quality adjusted price of~a~set of~products $G_t$ in month $t$ can be written as:

\begin{equation}
\tilde{p}^{t}=
\prod_{t \in G_{t}}\left(\frac{p_{i}^{t}}{v_{i}}\right)^{s_{t}^{t}}.
\end{equation}

Following \citet{diewert2004stochastic}, we assume that model (12) is estimated by the Weighted Least Squares (WLS) method with expenditure shares used as weights. Where the price is adjusted using the item fixed effects, i.e. $v_i=\exp(\gamma_i)$,  the TPD index can be expressed as follows:

\begin{equation}
P_{T P D}^{0, t}=
\frac{\tilde{p}^{t}}{
\tilde{p}^{0}}=
\frac{
\prod\limits_{i \in G_{t}}\left(\frac{p_{i}^{t}}{\exp \left(\hat{\gamma}_{i}\right)}\right)^{s_{i}^{t}}}{
\prod\limits_{i \in \sigma_{0}}\left(\frac{p_{i}^{0}}{\exp \left(\hat{\gamma}_{i}\right)}\right)^{s_{i}^{0}}}.
\end{equation}

\subsection{The updating problem and window updating methods}\label{sec-splicing-methods}

In the case of~bilateral methods, a~fixed base month (period) is used and the current period is shifted each month. In monthly chained index methods, the base and the current month are both moved one month. The problem with proceeding with the next month arises in the case of~multilateral index methods. Adding information from a~new month can influence the values of~quality adjustment parameters and values of~the corresponding multilateral indices. In this article, we consider four commonly used rolling-window updating methods, which shift the estimation window (often 13 months) forward in each period (a month as a~rule) and then splice the new indices onto the existing time series. The following methods are considered in our study.

\subsubsection{The movement splice method}

According to the movement splice method \citep{de2011eliminating}, a~price index for the new month is calculated by chaining the month-on-month index for the last month of~the shifted window to the index of~the previous month (the last month of~the previous window). The movement splice method can be described by the following recursive formula:

\begin{equation}
P_{M S}^{0, t}=P_{M S}^{0, t-1} \cdot P_{t-T, t}^{t-1, t}
\end{equation}

\noindent where $P$ is any multilateral price index formula, the subscript makes reference to the window period and the superscript indicates the period for which the index is calculated (see \citet{Loon2018}). 

\subsubsection{The window splice method}

The window splice method proposed by \citet{krsinich2014fews} is used to calculate the price index for the new month by chaining the indices of~the shifted window to the index of~$T$ months ago (i.e. to the index of~12 months ago for windows of~13 months). It can be expressed in the following general formula:

\begin{equation}
P_{W S}^{0, t}=P_{W S}^{0, t-1} 
\cdot 
\frac{P_{t-T, t}^{t-T, t}}{P_{t-T-1, t-1}^{t-T, t-1}}
\end{equation}

\subsubsection{The half splice method}

\citet{de2015framework} suggested that the link period $t_0$ should be chosen to be in the middle of~the first time window and the \citet{ABS2016} called this the half splice method for linking the results of~two time windows. In other words, according to this method, the half splice happens at $t_0 = (T+1)/2$ if $T$ is an odd integer and at $t_0=T/2$ if $T$ is an even integer. A recursive formula for the half splice method is as follows:

\begin{equation}
P_{H S}^{0, t}=P_{H S}^{0, t-1} \cdot \frac{P_{t-T, t}^{t-t_{0}, t}}{P_{t-T-1, t-1}^{t-t_{0}, t-1}}.
\end{equation}

\subsubsection{The mean splice method}

The mean splice method \citep{diewert2018substitution} uses the geometric mean of~all possible choices of~splicing, i.e. all months $\{1, 2, ..., T\}$ which are included in the current window and the previous one. The general formula for the mean splice method can be written as:

\begin{equation}
P_{G M S}^{0, t}=P_{G M S}^{0, t-1} \cdot \prod_{t_{0}=1}^{T}\left(\frac{P_{t-T, t}^{t-t_{0}, t}}{P_{t-T-1, t-1}^{t-t_{0}, t-1}}\right)^{\frac{1}{T}}.
\end{equation}  

\subsubsection{Other methods}

Instead of~using a~rolling
13-month window, \citet{chessa2016new} proposed a~method where “a time window is used with December of~the previous year as fixed base month. The first window consists of~the base month and January of~the present year”. The window expands every month to include the current month until December, when “the base month is shifted to December of~the current year” (\textit{Fixed Base Monthly Expanding Window} -- FBEW). \citet{lamboray2017geary} considered a~mix of~the FBEW method and the \textit{movement splice}. In his approach a~rolling window is used, where the last month of~the current period is compared to the previous December. In this case, December serves as the fixed base, as in the FBEW method. This method is called \textit{Fixed Base Moving Window} method (FBMW).

\subsection{Approaches to choosing a~price index for scanner data}

\begin{figure}[ht]
    \centering
    \includegraphics[width=0.7\textwidth]{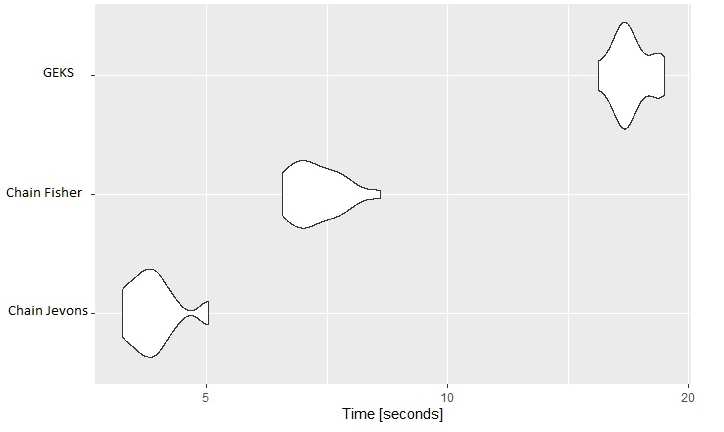}
    \caption{Comparison of~computation times for three price index methods (data set: milk, period: Dec. 2018 -- Dec. 2019)}
    \label{fig-speed-comp}
\end{figure}

Although \citet{ivancic2011scanner} have suggested that the use of~multilateral indices for scanner data can solve the chain drift problem, most statistical agencies that rely on scanner data still use the monthly chain Jevons index \citep{chessa2017comparison}. The choice of~the chain Jevons formula is quite understandable: a) after getting data from the current month, product matching is limited to the current and the previous month, i.e. generally, product matching is always made for two subsequent months; b) eventual data filtering and imputing is also limited to these subsequent months; c) quantities do not have to be imputed, while prices missing in the current month or flagged by filters have to be imputed based on historical prices or price changes; d) because of~a, b and c, the calculation of~the chained Jevons formula is much less time-consuming than calculating weighted formulas, in particular, those used in multilateral methods, which compare many pairs of~periods from the considered time window. To demonstrate the differences in the time it takes to determine various index formulas, let us use a~milk data set from one retail chain in Poland. The monthly data covered the period from December 2018 to December 2019. All necessary stages (product classification, product matching and data filtering) were carried out on the data set containing over 183,000 records. For demonstration purposes, the calculation times for the following index formulas are compared: the chain Jevons, the chain Fisher and the GEKS. The results for one hundred calculations carried out on one of~our servers (16 cores, 256 GB RAM) are presented in Figure \ref{fig-speed-comp}. As can be seen, the average computation time for multilateral formulas can be as much as four times longer than the time required to calculate unweighted bilateral formulas. If this difference is multiplied by the number of~outlets, the number of~retail chains and the number of~homogeneous product subgroups, it can be a~crucial factor from the perspective of~regular production of~price indices.

Multilateral indexes (and not only) are usually compared either using the axiomatic/test approach or the economic approach. In the first case, some desirable tests are defined that a~multilateral index can or cannot satisfy. The Australian Bureau of~Statistics has examined various tests for multilateral comparison (ABS, 2017).  For instance, the TPD and the Geary-Khamis indices do not pass the \textit{responsiveness test} and none of~the multilateral indices considered passes the \textit{identity test}. The Geary-Khamis index satisfies the \textit{basket test}, while the GEKS and the TPD indices do not; however, the Geary-Khamis formula does not pass the \textit{homogeneity in quantities}. The economic approach assumes that quantities are a~function of~prices but the usefulness of~this approach is limited in the case of~scanner data. One reason why this is the case is the fact that this approach does not adequately handle new and disappearing goods, which are regularly observed in the scanner data universe. Some authors compare multilateral methods in terms of~their \textit{flexibility} \citep{chessa2017comparison}. For instance, the GEKS, TPD and Geary-Khamis indices only use prices and quantities of~homogeneous products as input. Nevertheless, it is possible to extend these methods and explicitly incorporate additional characteristics of~products (if such variables have been included by data providers). For instance, product characteristics can be used as explanatory variables in the extension of~the regression model described in (12). To sum up, it can be concluded that the choice of~an optimal index formula for scanner data remains an open problem.

\section{Preparation of~scanner data sets}\label{prepare-data}

\subsection{Description of~scanner data sets used in the study}

Although empirical research on price indices concerns food products (section \ref{sec-comparison}), we have practised product classification and product matching on a~wider range of~products. First of~all, we wanted to verify our machine-learning methods by testing them on different products, and secondly, in the case of~food products, our time series were longer. Section \ref{prepare-data} presents one of~our preliminary studies conducted on a sample of product groups taken from both retail chains.

Scanner data used in the study were provided by two retail chains in Poland. Table \ref{tab-row-count} presents counts of~transactions (records in the database, not the number of~products purchased), which ranged from 5 million records in November 2017 to nearly 6.8 million records in October 2018. Each file was about 1~GB, totalling nearly 13~GB of~raw data.

\begin{table}[ht!]
    \centering
    \caption{Number of~records in the database}
    \label{tab-row-count}
    \centering
    \begin{tabular}{rr}
\hline
Year month & Number of~records\\
\hline 
2017/10 & 5 295 534  \\
2017/11 & 5 040 992  \\
2017/12 & 5 459 209  \\
2018/01 & 5 438 982  \\
2018/02 & 5 540 121  \\
2018/03 & 5 868 054  \\
2018/04 & 6 187 084  \\
2018/05 & 6 216 205  \\
2018/06 & 6 301 974  \\
2018/07 & 6 466 215  \\
2018/08 & 6 397 034  \\
2018/09 & 6 439 871  \\ 
2018/10 & 6 758 450  \\
\hline
    \end{tabular}
    \begin{flushleft}
     \footnotesize Note: numbers in the table represent transactions (records) and not products purchased. The data provided by retail chains were aggregated, i.e. each record represented a~specific transaction (or grouped transactions at a~specified price), which may have included many identical products.
    \end{flushleft}
   
\end{table}

One of~the main problems encountered at this stage was standardizing product descriptions to create a~set of~unique names that could be used to match products between chains and over time. The data provided by retail chains had not been cleaned and contained errors and non-homogeneous names. As a~result, name versions, such as "Butter" and "butter" (letter case), or "5L drink" and "5Ldrink" (spacing) would be treated as unique. The original set of~raw scanner data, containing 25~636 unique product names was sent to the Statistical Office in Opole, which has a~special team of~employees tasked with manual classification of~products. Our approach was to apply machine learning to classify products into COICOP groups based on their descriptions.

\subsection{Data preparation for product classification}\label{prod-class}

After the manual classification of~raw data, performed by the Statistical Office in Opole, there were 311 products names for which unambiguous COICOP groups could not be assigned. They were mainly products with very generic names (for example, BEER, DELICIOUS VEGETABLES) or ambiguous names (for example, UDA STEAK WITH TURKEY IN MARINATE). Table \ref{tab-grupy-proj} presents the number of~unique product names in selected COICOP groups.

\begin{table}[ht!]
\centering
\caption{Number of~unique product names in selected COICOP groups} 
\label{tab-grupy-proj}
\centering
\begin{tabular}{llrr}
  \hline
 COICOP code & COICOP name & Number of~product names & Share \\ 
  \hline
  11411 & Fresh whole milk & 68 & 0.29 \\ 
  11421 & Fresh low-fat milk & 99 & 0.43 \\ 
  11431 & Condensed and powdered milk & 32 & 0.16 \\ 
  11511 & Butter & 52 & 0.22 \\ 
  121321 & Toothpastes & 113 & 0.49 \\
  -- & Other & 22~868 & 98.4 \\ 
   \hline
\end{tabular}
\end{table}

Word processing (\textit{text mining}) was conducted using regular expressions (\textit{regular expression}, in short \textit{regex} or \textit{regexp}), i.e. patterns describing strings of~symbols. Regular expressions can specify a~set of~matching strings or relevant parts of a string. In order to classify products into COICOP groups, regular expressions were used to (1) extract measurement units (e.g. weight, volume) from product descriptions to new columns, (2) remove special characters and some abbreviations and (3) separate camelCase expressions (for example \textit{WodaZrodlanaPromocaj5L}).

\subsection{Product classification}\label{product-class-algo}

In the first step, we took into account the unbalanced nature of~the raw data, i.e. the fact that COICOP groups varied in terms of~the number or products. The procedure was as follows. From the complete set, containing deduplicated  product names classified into COICOP groups, 12 representatives of~each group were drawn (including the category \textit{Other}) to create a~test dataset that was not used in the learning process. The remaining items were included in a~training set according to the following procedure:

\begin{itemize}
               \item a~separate dataset was created for each category containing the remaining observations for each category and 400 observations from the category \textit{Other},
                \item the \textit{Synthetic Minority Over-sampling Technique}, developed by  \citet{chawla2002smote}, was applied to each category to create \textit{synthetic observations} for minority categories.
\end{itemize}
    
The resulting dataset contained 3~847 observations and~1~680 columns (unique strings). Next, we applied LASSO regression.
    
LASSO regression was proposed by \citet{santosa1986linear, tibshirani1996regression} and further developed by \citet{friedman2010regularization, tibshirani2012strong, hastie2015statistical, zou2006adaptive}.  The main idea behind LASSO regression is to minimize the following likelihood (loss) function:

\begin{equation}
\min_{\beta_0,\beta} \frac{1}{N} \sum_{i=1}^{N} w_i l(y_i,\beta_0+\beta^T x_i) + \lambda\left[(1-\alpha)||\beta||_2^2/2 + \alpha ||\beta||_1\right],
\end{equation}

\noindent where $l()$ is the negative log-likelihood function, $\beta_{0}, \beta$ are model parameters, $x_i$ are independent variables, and  $w$ are case weights. If $\alpha=1$, we get LASSO regression, and if $\alpha=0$, we get ridge regression. 

Because our target variable is polytomous, we considered multinomial LASSO with $K$ levels with set ${\cal G}=\{1,2,\ldots,K\}$ and the multinomial logistic regression model given by the following equation:

\begin{equation}
\mbox{Pr}(G=k|X=x)=\frac{e^{\beta_{0k}+\beta_k^Tx}}{\sum_{\ell=1}^Ke^{\beta_{0\ell}+\beta_\ell^Tx}}.
\end{equation}
    
Let $\bY$ be an matrix of~size $N \times K$, where elements are defined by $y_{i\ell} = I(g_i=\ell)$. Then the log-likelihood (loss) function is given by: 

\begin{equation}
    \begin{split}
        \ell(\{\beta_{0k},\beta_{k}\}_1^K) &= -\left[\frac{1}{N} \sum_{i=1}^N \Big(\sum_{k=1}^Ky_{il} (\beta_{0k} + x_i^T \beta_k) - \log \big(\sum_{k=1}^K e^{\beta_{0k}+x_i^T \beta_k}\big)\Big)\right] \\ 
        &+\lambda \left[ (1-\alpha)||\beta||_F^2/2 + \alpha\sum_{j=1}^p||\beta_j||_q\right],
    \end{split}
    \end{equation}
    
\noindent where $\bbeta$ is a~matrix of~model parameters with $p \times K$ dimensions, $\bbeta_k$ refers to $k$-th level of~$\bY$, and~$\bbeta_j$ refers to $j$-th row ($K$ vector of~$j$-th variable).  

Model selection was based on 10-fold cross-validation and minimization of~misclassification error. During the procedure, the $\lambda$ parameter was optimised, resulting in a~set of~parameters that meet the above-mentioned criterion. We used R statistical software \citep{rcran} and the \texttt{glmnet} package \citep{glmnet1,glmnet2}.

Table \ref{tab-uczenie-uczacy} contains results of~the training and testing procedures. The columns refer to predicted COICOP codes and rows to true (observed) codes. As can be seen, the prediction of~COICOP groups based on product descriptions using LASSO regression was nearly perfect, with only one product incorrectly classified, (as fresh low-fat milk instead of~Fresh whole milk). Prediction results for the test dataset, which was not used for training, indicate that this algorithm has a~strong predictive power, as only one product was wrongly classified. 

\begin{table}[ht]
\centering
\caption{Results of~training and testing with 10-fold CV using multinomial LASSO regression}
\label{tab-uczenie-uczacy}
\begin{center}
\begin{tabular}{rrrrrrr}
  \hline
True/Predicted & 11411 & 11421 & 11431 & 11511 & 121321 & Other \\ 
  \hline
  \multicolumn{7}{c}{Training dataset} \\
  \hline
  11411 -- Fresh whole milk & 391 &   1 &   0 &   0 &   0 &   0 \\ 
  11421 -- Fresh low-fat milk &   0 & 352 &   0 &   0 &   0 &   0 \\ 
  11431 -- Condensed and powdered milk &   0 &   0 & 400 &   0 &   0 &   0 \\ 
  11511 -- Butter &   0 &   0 &   0 & 400 &   0 &   0 \\ 
  121321 -- Toothpastes &   0 &   0 &   0 &   0 & 303 &   0 \\ 
  Other &   0 &   0 &   0 &   0 &   0 & 2000 \\
   \hline
   \multicolumn{7}{c}{Test dataset} \\
   \hline
  11411 -- Fresh whole milk &  12 &   0 &   0 &   0 &   0 &   0 \\ 
  11421 -- Fresh low-fat milk &   0 &  11 &   1 &   0 &   0 &   0 \\ 
  11431 -- Condensed and powdered milk &   0 &   0 &  12 &   0 &   0 &   0 \\ 
  11511 -- Butter &   0 &   0 &   0 &  12 &   0 &   0 \\ 
  121321 -- Toothpastes &   0 &   0 &   0 &   0 &  12 &   0 \\
  Other &   0 &   0 &   0 &   0 &   0 &  12 \\ 
  \hline
\end{tabular}
\end{center}
\end{table}

Finally, to verify how good LASSO regression was at predicting new cases, we used 22~236 product names in the “Other” category (including those used in the training dataset). Table \ref{tab-uczenie-pozost} presents prediction results for this group of~products. 99.26\% of~all observations were correctly classified as “Other” and only 164 were misclassified into one of~the five remaining COICOP categories. These products were mainly baby milk powder, flavored milk, buttermilk, sandwich spreads or toothbrushes.

\begin{table}[ht]
\centering
\caption{Prediction results for products classified as Other}
\label{tab-uczenie-pozost}
\begin{center}
\begin{tabular}{llrr}
  \hline
True COICOP & Predicted COICOP  & Count & Share \\ 
  \hline
 Other & 11411 -- Fresh whole milk &   1 & 0.00 \\ 
 Other & 11421 -- Fresh low-fat milk &  15 & 0.07 \\ 
 Other & 11431 -- Condensed and powdered milk &  20 & 0.09 \\ 
 Other & 11511 -- Butter &  29 & 0.13 \\ 
 Other & 121321 -- Toothpastes &  99 & 0.45 \\ 
 Other & Other & 22 069 & 99.26 \\ 
   \hline
\end{tabular}
\end{center}
\end{table}

\subsection{Product matching – theoretical assumptions and results}\label{sec-matching}

In this section, we briefly describe the matching procedure applied in points of sale of~a given data provider. Most countries that apply the Jevons chain index for scanner data use matching only for two months (periods), the current one and the one immediately preceding. However, when multilateral indices are used, products are matched over the entire time window for all compared pairs of~months. This procedure is more time-consuming than in the case of~chained indices, where products are matched “only” for adjacent months (periods), i.e. $t-1$ and $t$. The following steps can be identified in the matching procedure:

\begin{enumerate}
    \item \textit{matching by identifier} -- EAN code or provider ID can be used; 
    \item \textit{matching by name and blocking} -- when products have the same provider ID but different EAN codes (i.e. due to changes in the products, such as packaging color), two products are regarded as matched if they have the same or similar name (based on text dissimilarity measures, see for instance \citet{strdist} for overview). If products have different EAN codes and provider IDs, matching occurs when products have exactly the same description (label). Blocking is applied to groups of~products with exactly the same characteristics (e.g. weight);
    \item \textit{matching by name with limited blocking} -- products with different codes and descriptions which are, in fact, the same from the customer's point of~view are considered as matched (e.g. different weight and packaging, resulting in a~different label and description; \textit{relaunched products}) . 
\end{enumerate}

In our study we used the following two-step procedure for matching products over time:

\begin{enumerate}
    \item Step 1 -- matching using EAN codes and product IDs,
    \item Step 2 -- for products not matched in Step 1 apply matching with blocking.
\end{enumerate}

The second step involved probabilistic record linkage using the \textit{reclin} R package \citep{reclin} with blocking by provider ID, point of sale ID, percentage value (if available, e.g. milk), and whether milk is UHT (Boolean variable), weight and measurement unit. We allowed descriptions to differ and used the Jaro-Winkler similarity given by:

\begin{equation}
\operatorname{sim}_{j}=\left\{\begin{array}{ll}{0} & {\text { if } m=0}, \\ {\frac{1}{3}\left(\frac{m}{\left|s_{1}\right|}+\frac{m}{\left|s_{2}\right|}+\frac{m-t}{m}\right)} & {\text { otherwise, }}\end{array}\right.
\label{jaro}
\end{equation}

\noindent where $\left|s_{i}\right|$ is the length of~string $s_i$, $m$ denotes the number of~exactly the same characters, and $t$ is half of~the number of~transpositions required to obtain the same string.  In the study we used a~similarity score given by equation \eqref{jaro-score}
        
\begin{equation}
    \text{sim}_w = \text{sim}_j + l\cdot p\cdot(1-\text{sim}_j),
    \label{jaro-score}
\end{equation}

\noindent where $l$ denotes the number of~leading characters (a prefix of~up to 4 characters), and $p$ is a~scaling factor set to 0.1, following Winkler. 

\begin{figure}[ht!]
    \centering
    \includegraphics[width=\textwidth]{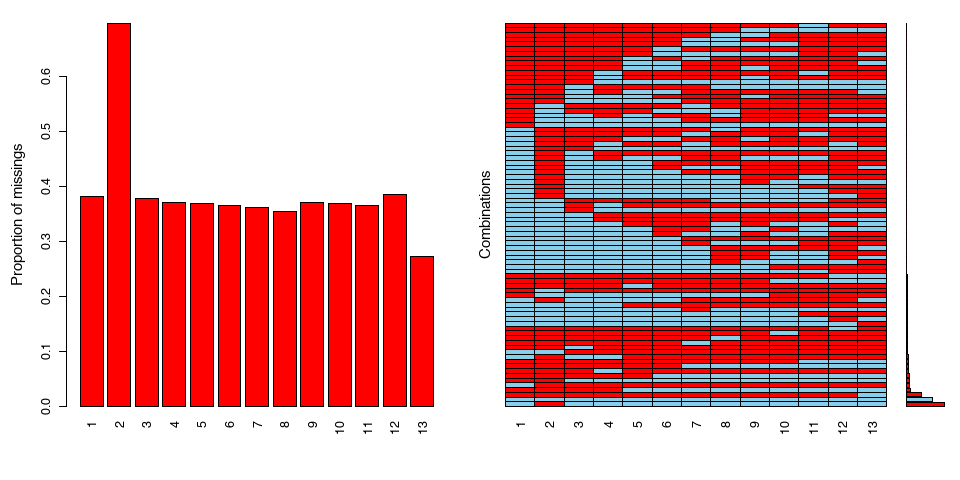}
    \caption{Distribution of~products over time. The left plot shows the proportion of~missing observations in the period from Nov 2018 to Dec 2019. The right plot shows combinations of~interruptions in product availability. Missing data are marked in red, while months when data were available are marked in blue.}
    \label{fig-matched}
\end{figure}

The final database consisted of~304 products, compared to 352 product before matching and 312 products matched solely using provider IDs. Figure \ref{fig-matched} presents the distribution of~matched products over the reference period.

\section{Data filtering  and its influence on price index values}\label{sec-filtering}

There is an ongoing discussion in the literature about whether or not to use data filters for scanner data, and if so, what kind of~filters to apply. As a~rule, scanner data indices are calculated using a~dynamic approach, with most countries opting for the~monthly chain Jevons index. This method is commonly referred to as the dynamic method \citep{eurostat2017}. The dynamic basket is determined using turnover figures of~individual products in two adjacent months, i.e. the product is included in the sample if its turnover is above a~fixed threshold determined by the number of~products in a given product group. \citet{Loon2018} provide the following condition for the above mentioned rule, which indicates whether the $i$-th product is taken into consideration in the comparison of~months $t-1$ and $t$:

\begin{equation}
    \frac{s_i^{t-1} + s_{i}^{t}}{2} > \frac{1}{n\lambda},
\end{equation}

\noindent where $n$ is the number of~considered products and $\lambda$ is a~fixed parameter (usually set to 1.25). This kind of~data filter can be called a~\textit{low sale filter}. Supporters of~using filters also believe that products displaying extreme price changes from one month to another should also be excluded from the sample (\textit{extreme price filter}). The list of~possible data filters is longer, e.g. Statistics Belgium implements a~filter for dump prices \citep{Loon2018}. 

\begin{figure}[ht!]
    \centering
    \includegraphics[width=0.45\textwidth]{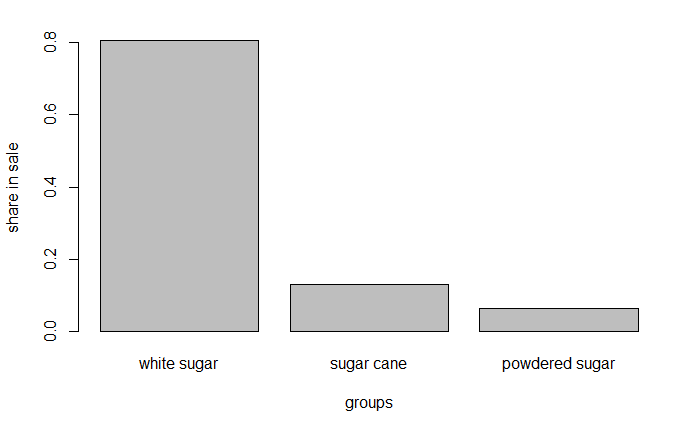}
    \includegraphics[width=0.45\textwidth]{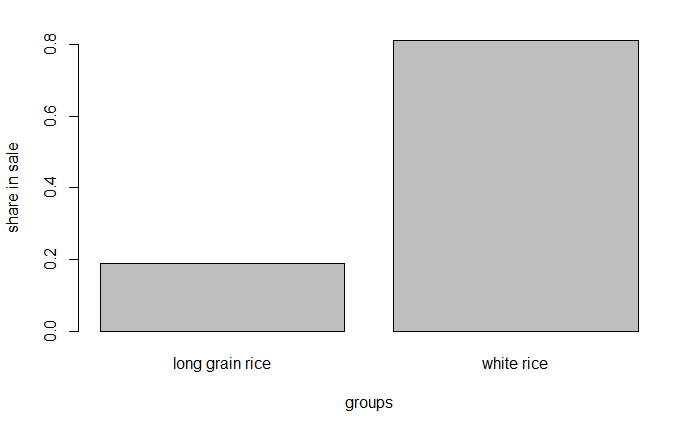}
    \caption{Shares of~sales for subgroups of~sugar and rice (Dec 2018 -- Dec 2019: without filtering)}
    \label{fig-shares-init}
\end{figure}

\begin{figure}[ht!]
    \centering
    \includegraphics[width=0.45\textwidth]{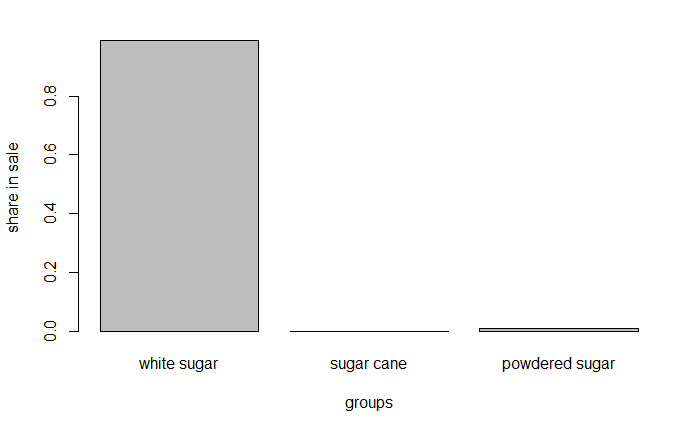}
    \includegraphics[width=0.45\textwidth]{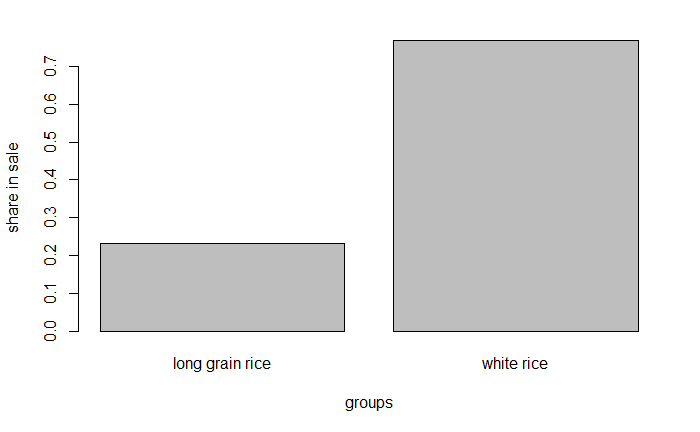}
    \caption{Shares of~sales for subgroups of~sugar and rice (Dec 2018 -- Dec 2019: after LS2 filtering)}
    \label{fig-shares-filtered}
\end{figure}

Filtering products is one thing and deciding what to about products that have been filtered out is another. One possible option is to impute prices flagged by filters, but this raises questions about the method of~imputation. The second option is to remove flagged products from the sample if it does not change the sample size critically. In this section we examine the impact of~using the \textit{low sale filter} and the \textit{extreme price filter} on price index values under this second option. In the empirical example we use scanner data from one retail chain in Poland, i.e. monthly data from over 210 outlets on \textit{sugar} and \textit{rice} (together with their more homogeneous subgroups below the COICOP 5 level) sold between December 2018 and December 2019. We consider the following filter variants: \textit{low sale filters} with $\lambda=1$ (LS1), $\lambda=1.25$ (LS2) and $\lambda=1.5$ (LS3), \textit{extreme price filters} with thresholds for the minimum and maximum price change set to 50\%  and 200\% (EP1), and also 25\% and 300\% (EP2), respectively. We also include another variant of~the \textit{extreme price filter}, namely the filter which removes products with price changes (in compared months) that are smaller and bigger than the $1^{\text{th}}$  and $99^{\text{th}}$ quantile of~all observed price changes (EP3). First, we determine structures of~expenditure shares in the analysed product groups, taking into account their homogeneous subgroups (homogeneous in terms of~the local level of~COCIOP 6). These structures for the whole 13-month time interval are presented in Figure \ref{fig-shares-init}. Figure \ref{fig-shares-filtered} presents analogous structures after using the most typical low sale filter (LS2).  

\begin{figure}[ht!]
    \centering
    \includegraphics[width=0.45\textwidth]{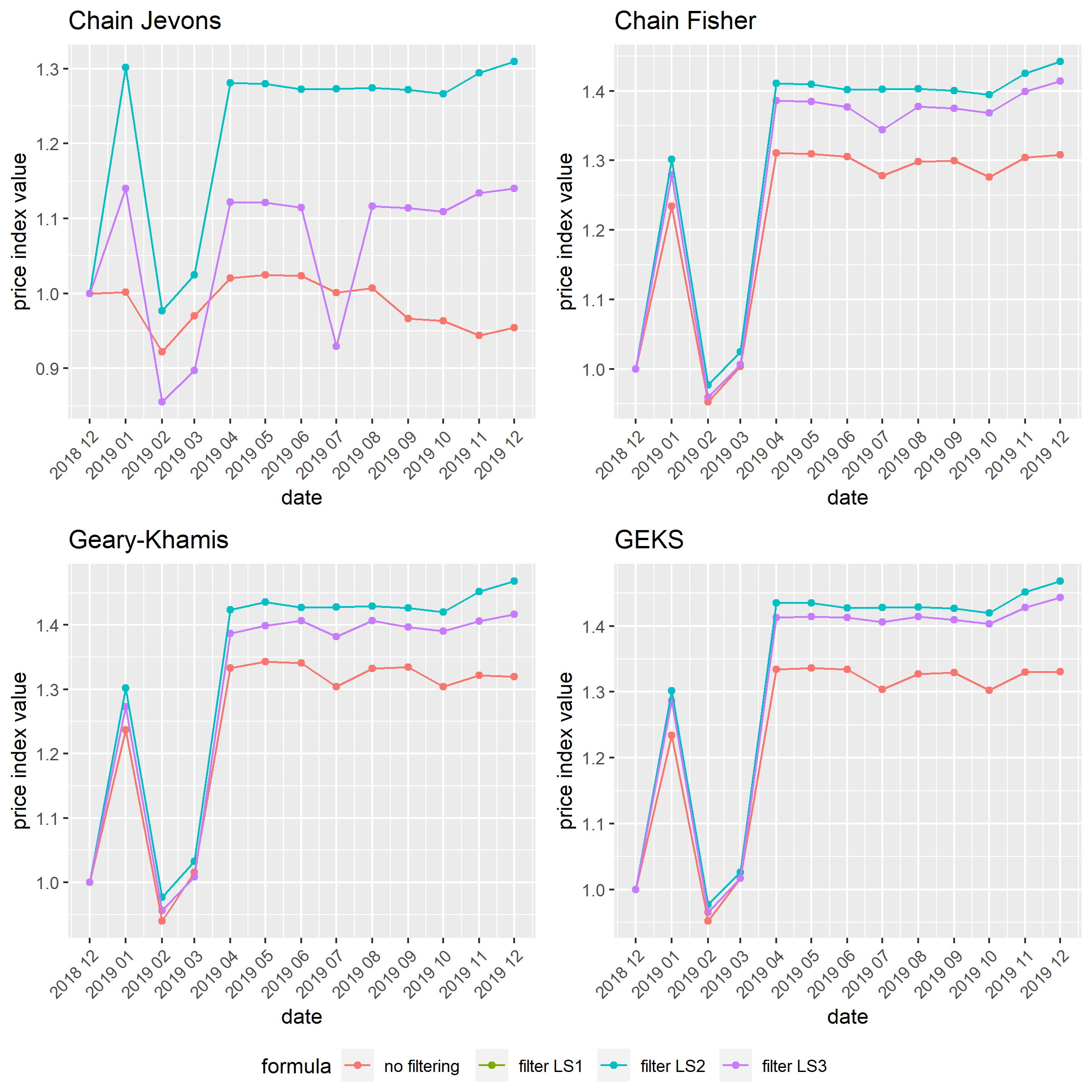}  
    \includegraphics[width=0.45\textwidth]{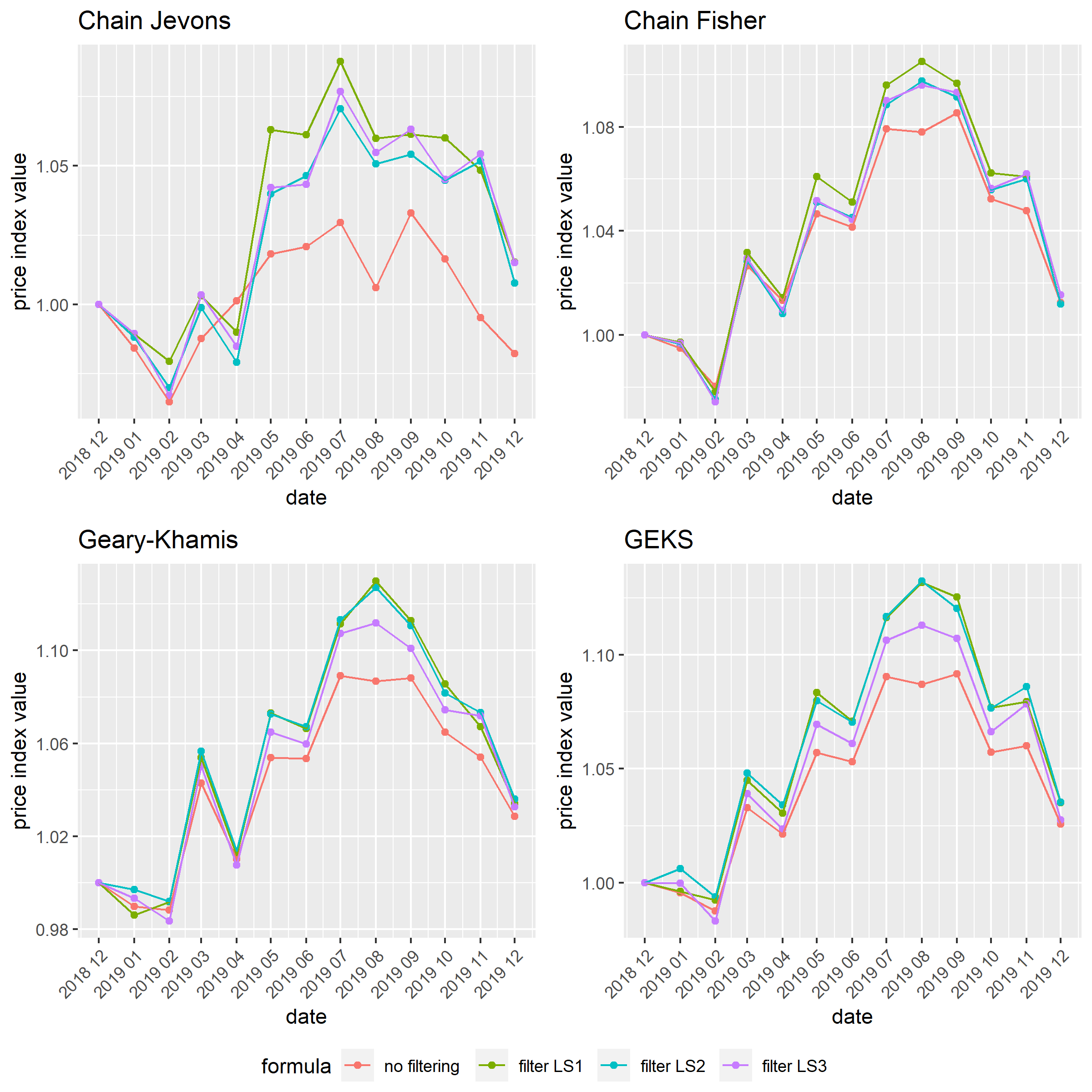}  
    \caption{The effect of~low sale filters on price index values}
    \label{fig-price-low}
\end{figure}

\begin{figure}[ht!]
    \centering
    \includegraphics[width=0.45\textwidth]{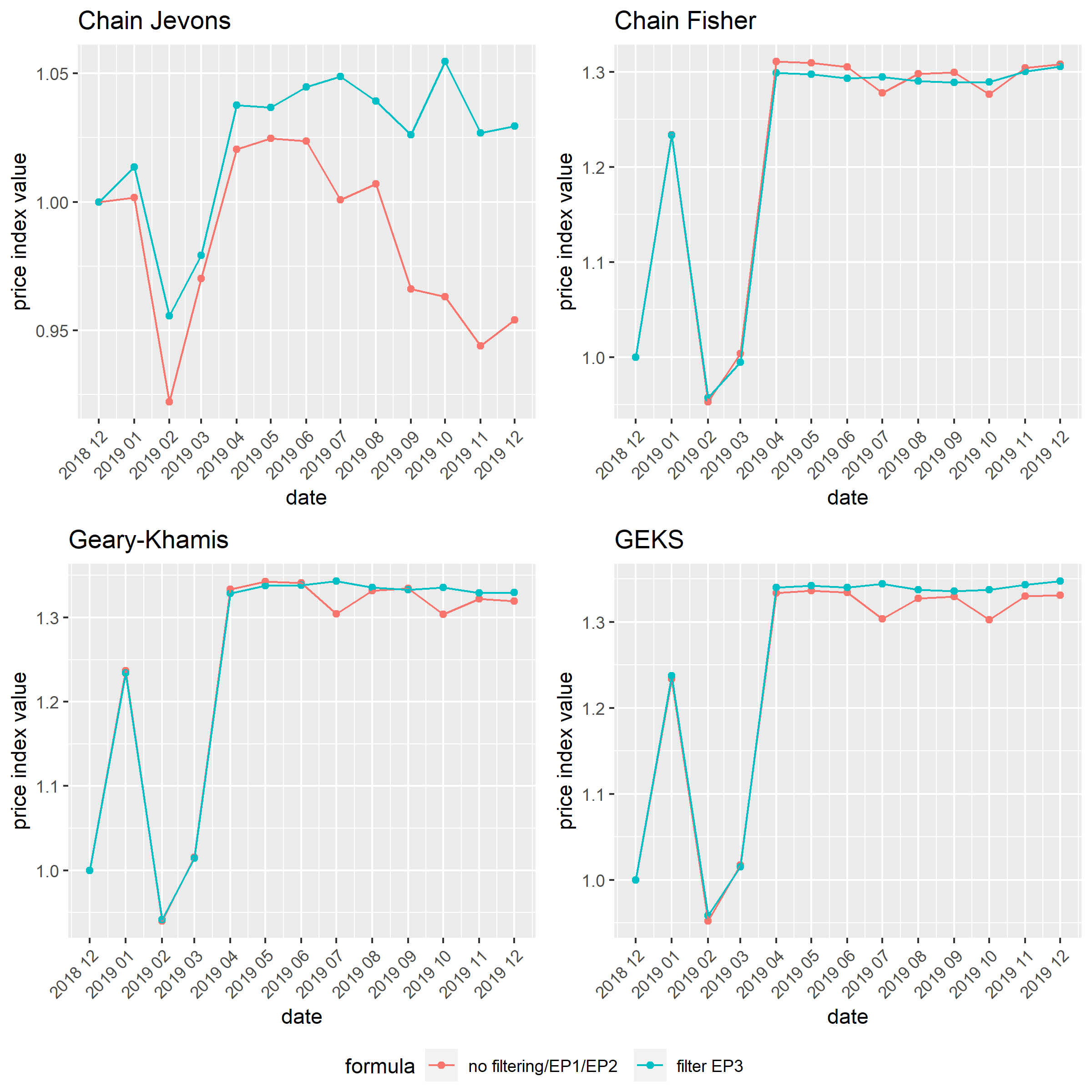}
    \includegraphics[width=0.45\textwidth]{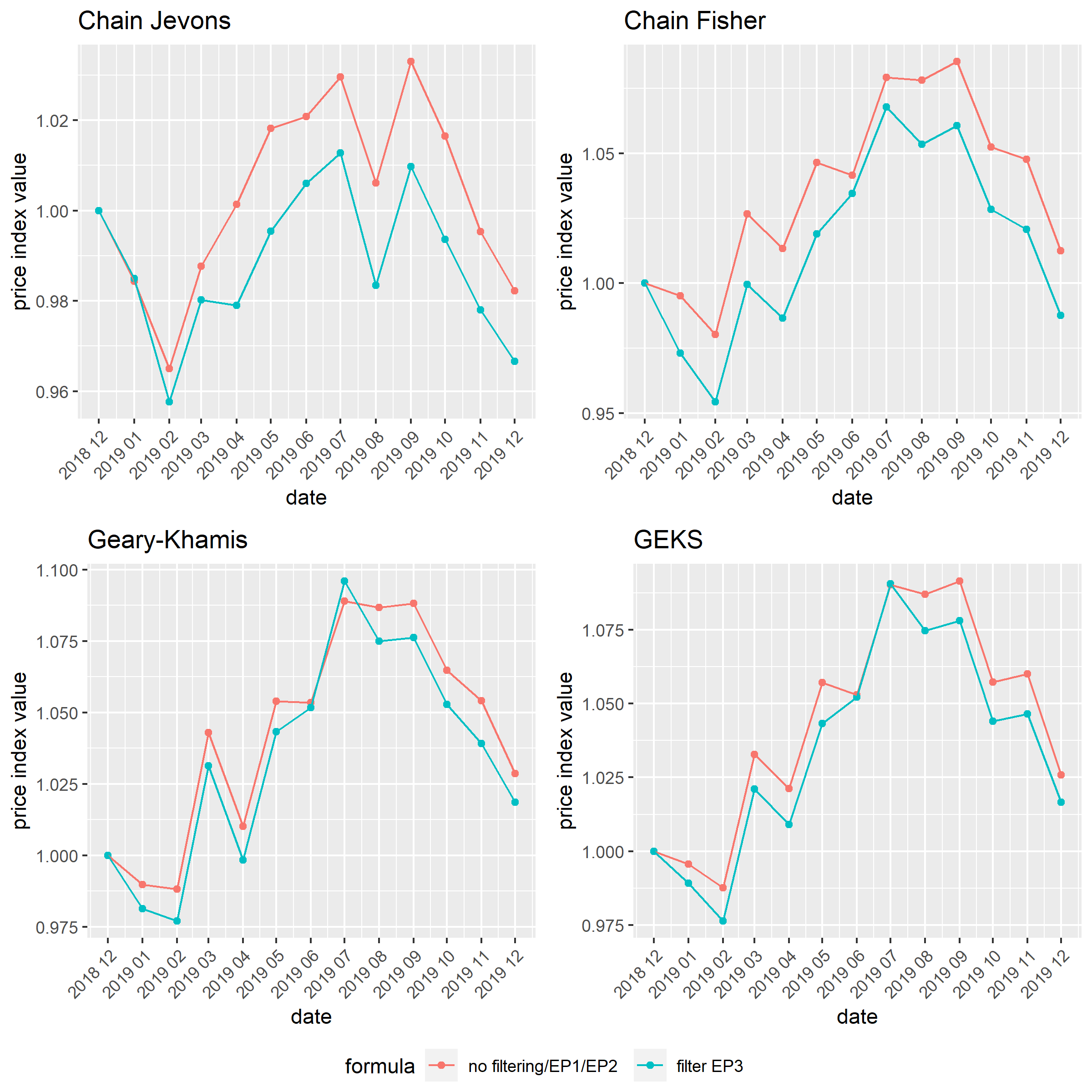}
    \caption{The effect of~extreme sale filters on price index values}
    \label{fig-price-ex}
\end{figure}

Note that after applying the filter the structure of~sales has changed, in the case of~sugar considerably . Therefore, the type of~filter used and its threshold values can affect the measured price dynamics. This possibility is verified by using the chain Jevons, the chain Fisher, the full-window Geary-Khamis and GEKS indices (see Fig. \ref{fig-price-low} and Fig. \ref{fig-price-ex}) with Dec. 2018 as the fixed base.

Based on Figures \ref{fig-price-low} and \ref{fig-price-ex}, it can be concluded that for both types of~filters the unweighted Jevons formula is the most sensitive to the type of~data filter and its thresholds. 

In the case of~\textit{low sale filters}, as a~rule, the LS1 filter generates the highest index values regardless of~the index formula and product group (see Fig. \ref{fig-price-low}). At the same time, the absence of~low sale filtering almost always leads to lower values of~price indices regardless of~the product group. Intuitively, this can be explained as follows: in the case of~products reduced to clear, price decreases are associated with a~lower demand for these products. Consequently, the value of~their sales also decreases. Therefore, if such products are not filtered out, some falling prices will be included and the index will ultimately be smaller. 

\begin{figure}[ht!]
    \centering
    \includegraphics[width=0.45\textwidth]{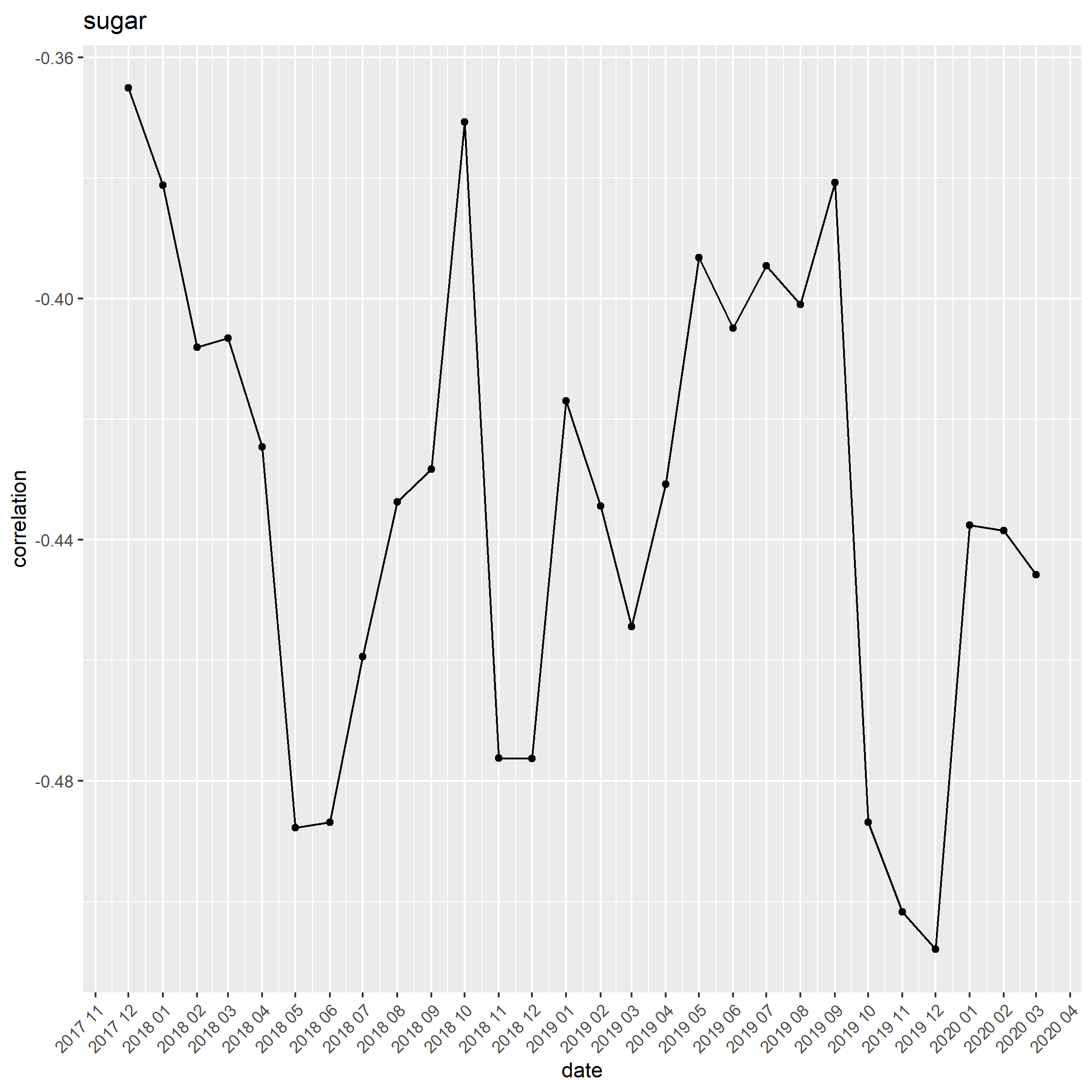}
    \includegraphics[width=0.45\textwidth]{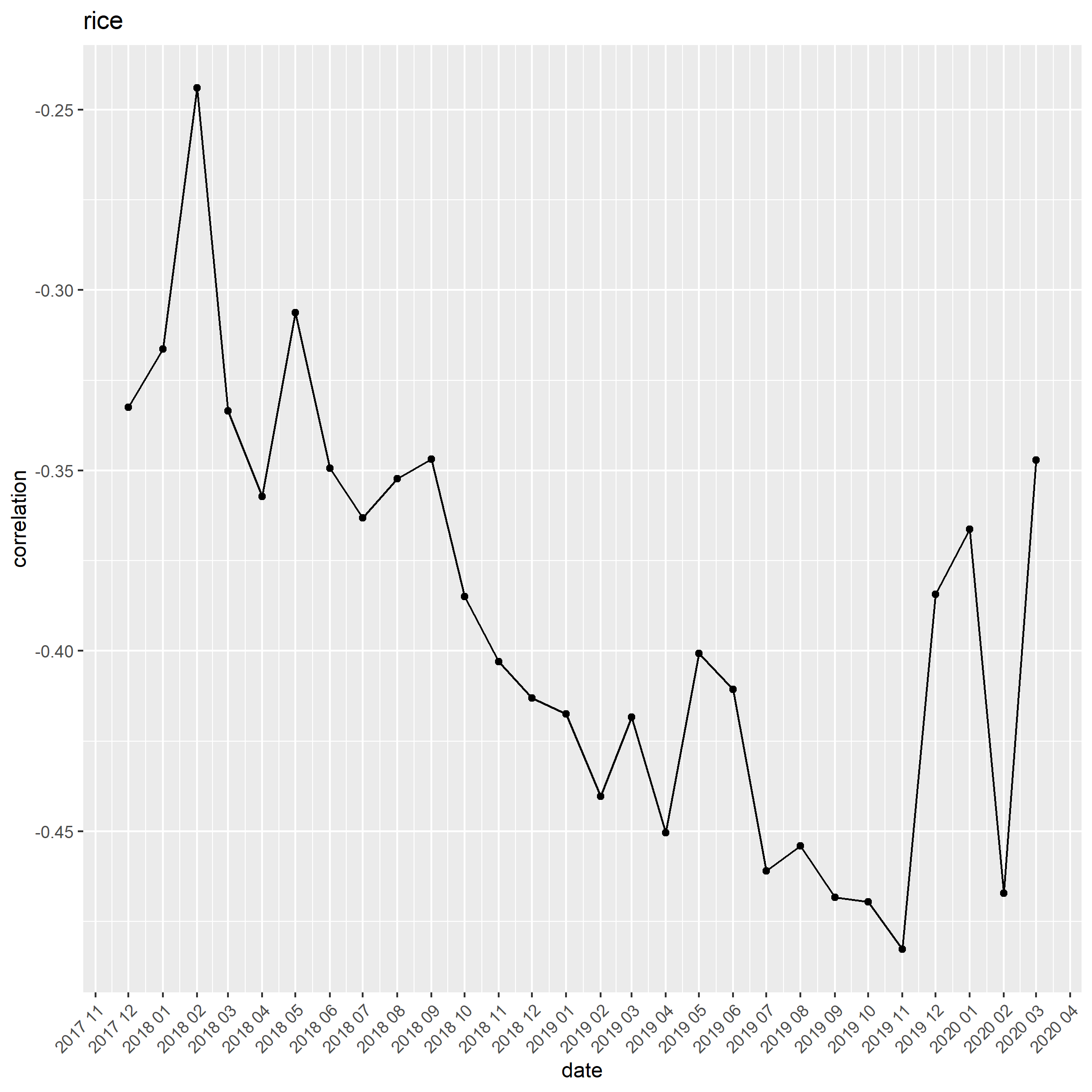}
    \caption{Monthly values of~Pearson's correlation coefficient showing the relationship between prices and quantities of~the studied products}
    \label{fig-corrs}
\end{figure}

In the case of~\textit{extreme price filters}, the choice between no filtering and the use of~EP1 and EP2 filters is not important in our study. This is because no price shocks were observed in the reference period for sugar and milk in the retail chain. Of course, the use of~the EP3 filter generates a~certain effect because it is based on the quantiles of~the distribution of~price changes, and not on predetermined thresholds. As with the previous filter type, the differences between the price index values with and without EP3 filters are much smaller in the case of~weighted index formulas (including GEKS) compared to the unweighted Jevons index (in this case differences may exceed 10 p.p. -- see Fig. \ref{fig-price-ex}). Interestingly, no definite conclusions can be drawn about the direction of~differences between the indices, i.e. the EP3 filter applied to rice generates lower index values than those obtained in the absence of~filtering; in the case of~sugar, the filter generates higher values (see Fig. \ref{fig-price-ex}). Intuition suggests that this may be related to the correlation between the price and demand for these products; in other words, it may depend on how consumers react to rising prices within these product groups. Nevertheless, Figure \ref{fig-corrs} seems to contradict this conjecture, because in both cases there is a~similar level of~correlation between prices and quantities determined for all products sold in subsequent months.

\section{Comparison of~price index methods: empirical study}\label{sec-comparison}

In the following empirical study we use scanner data from one retail chain in Poland, i.e. monthly data on \textit{milk}, \textit{sugar}, \textit{coffee} and \textit{rice} sold in over 210 outlets during the period from December 2017 to December 2019. We investigate several crucial problems connected with the choice of~the optimal price index formula for scanner data.  Products classified into COICOP 5 groups are matched using the \texttt{reclin} R package (as described in Section \ref{sec-matching}) and the standard LS2 filter is used to exclude products with relatively low sales. 

Note that depending on the type of~analysis, data source or index formula, the time interval used in the analysis can vary. Unfortunately, scanner data were delivered on a~regular and uninterrupted basis by only one of~the two retailed chains we cooperated with. However, our initial experiments with classification and matching were based on consistent data from both chains, hence, e.g. Tables \ref{tab-row-count}--\ref{tab-uczenie-pozost} relate to the period from October 2017 to October 2018. The experience gained in this initial phase was used to compile dictionaries of~words and phrases to identify products and to select machine-learning methods for automatic classification and matching in the following months. Because, as already mentioned, only one retail chain delivered data continuously, our study of index formulas was limited to scanner data from that one source. However, depending on the index formulas used or the type of~analysis, the reference period is not always the same, e.g. the FBMW method requires a~longer interval of~observations than the FBEW method; similarly, chain indices can be compared over a~period of~one year, whereas the time interval required to compare splicing indices with a~13-month time window should be longer than a~year. Sometimes, to facilitate the presentation, the time interval  was shortened owing to the time consuming nature of~some procedures. For example, given the capacity of our server, it could take several hours to calculate any splicing multilateral index with double aggregation across outlets and subgroups for the entire available period (over 30 months). Therefore, the presented figures with comparisons of~price index formulas do not always relate to the same period of~time. Similarly, the product groups discussed in section \ref{scanner-data} do not completely coincide with the product groups that were chosen in section \ref{sec-comparison}. More specifically, the product groups in sections \ref{prod-class} and \ref{sec-matching} were selected to make it easier for the reader to grasp the idea behind the methods of~classifying and matching products. However, in section \ref{sec-comparison}, which presents results of~the empirical study of~price indices, we selected product groups that could help us capture certain patterns in relations between the indices. As a~result, the set of~product groups considered in the article is wider and more diverse.

\subsection{Unweighted vs weighted formulas}\label{sec-weighted}

\begin{figure}[ht!]
    \centering
    \includegraphics[width=0.45\textwidth]{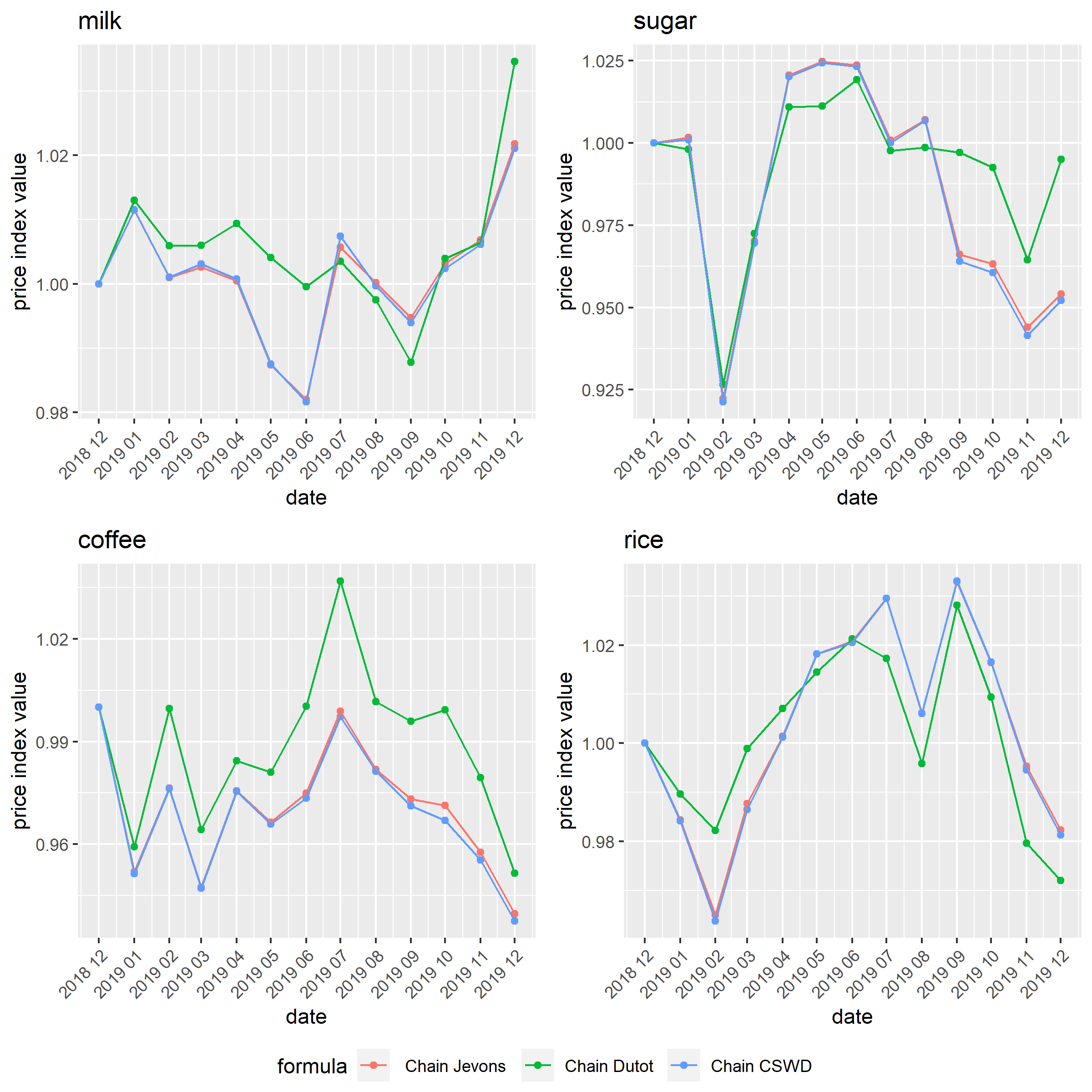}
    \includegraphics[width=0.45\textwidth]{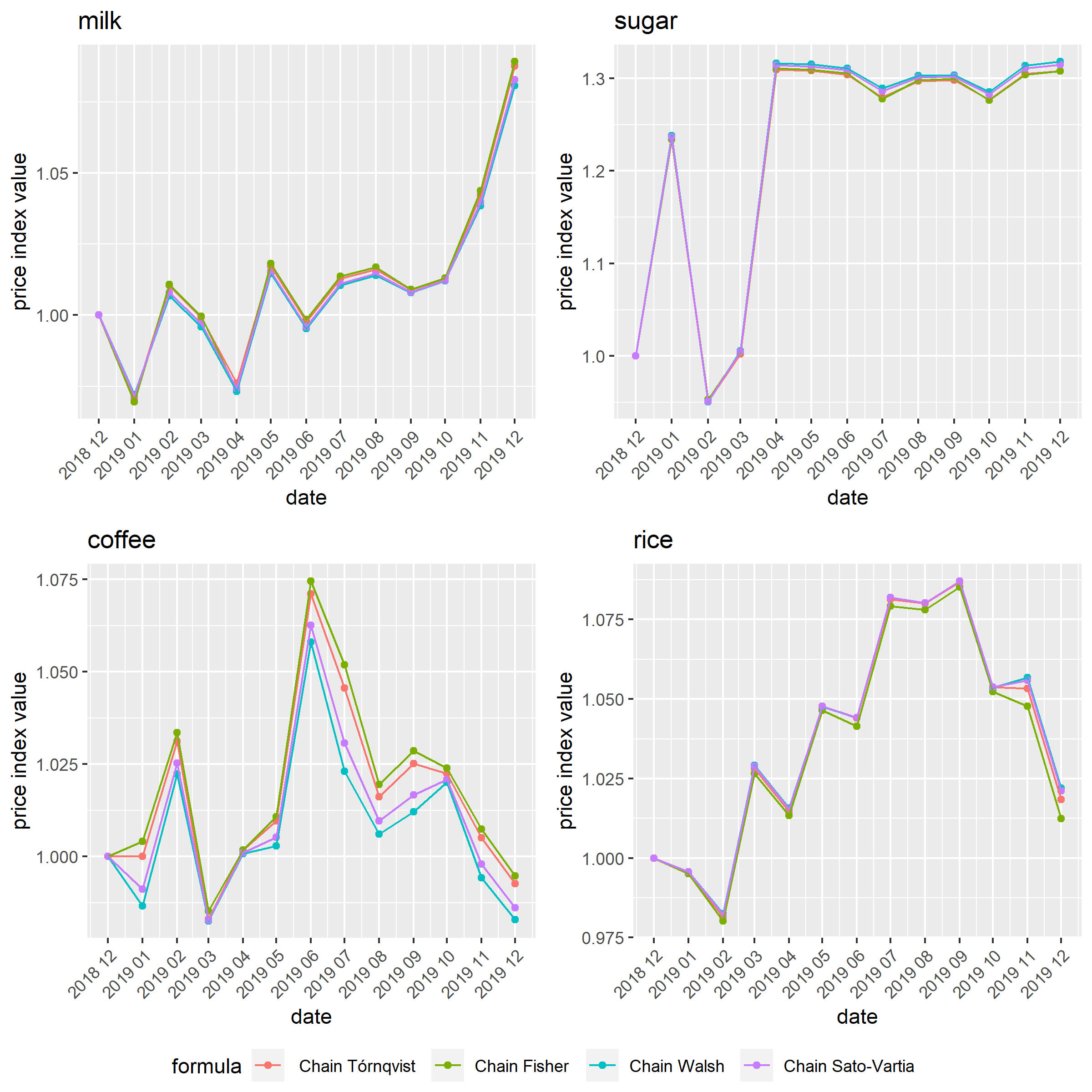}
    \caption{Comparison of~selected chain elementary price indices (two left columns), Comparison of~selected chain weighted price indices (two right columns)}
    \label{fig-chains-two}
\end{figure}

\begin{figure}[ht!]
    \centering
    \includegraphics[width=0.9\textwidth]{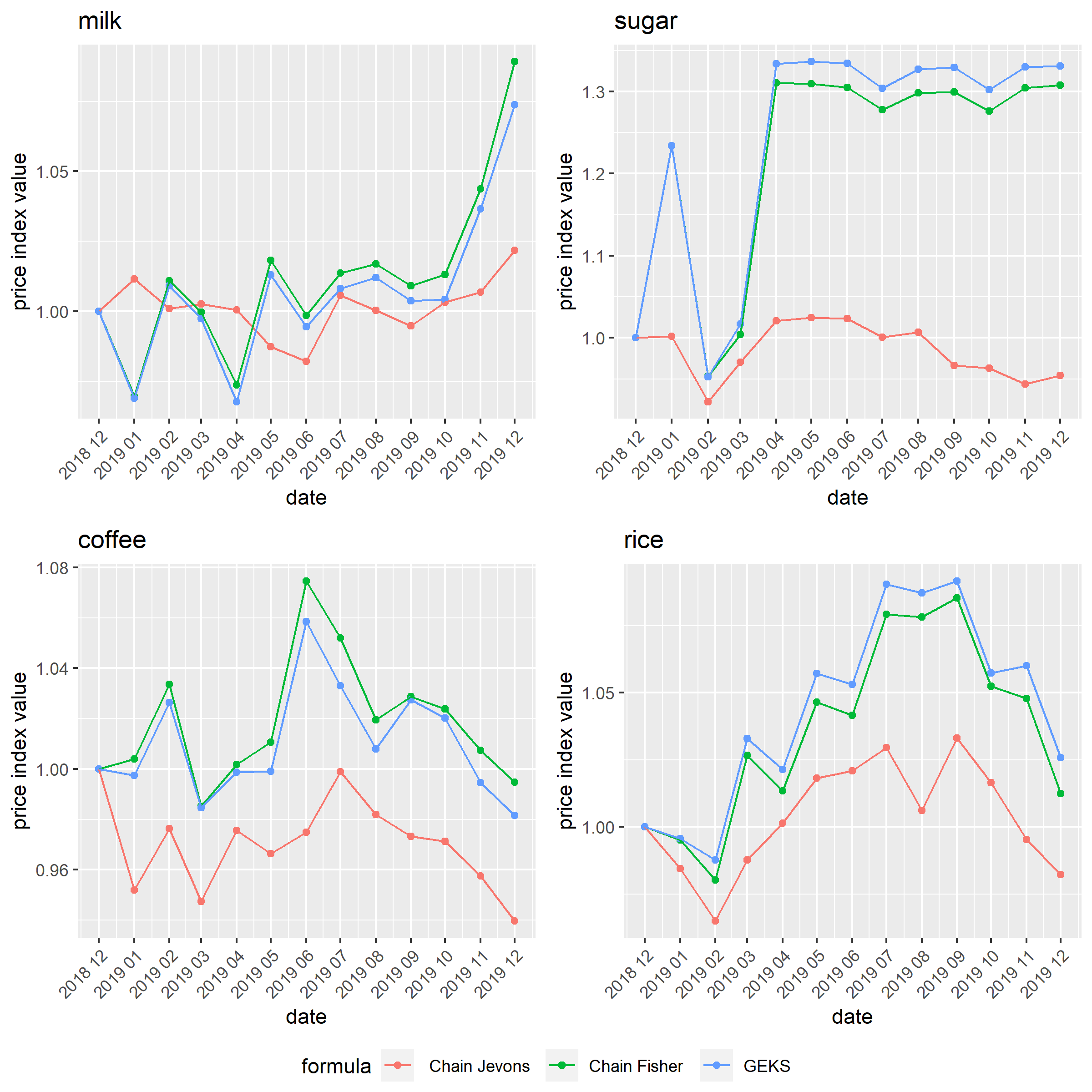}
    \caption{Comparison of~selected unweighted and weighted price index formulas}
    \label{fig-comp-weight}
\end{figure}

It is commonly known that the use of~multilateral indices for scanner data can solve the chain drift problem, but most statistical agencies still make use of~the monthly chain Jevons index for scanner data sets \citep{chessa2017comparisons}. Some NSIs use the chain Fisher or T\"ornqvist indices (e.g. USA or Japan), which are weighted formulas and take into consideration additional and available information about expenditures. As already mentioned, multilateral indices seem to be the right choice owing to their transitivity and the dynamic character of~scanner data sets. Figure \ref{fig-chains-two} (left) presents a~comparison of~selected chain elementary price indices calculated for the group of~products analysed in the study. Figure \ref{fig-chains-two} (right) compares selected chain weighted prices indices, including the chain Fisher formula. Figure \ref{fig-comp-weight} presents a~comparison between values of~the chain Jevons, the chain Fisher and the GEKS indices, which are commonly used with scanner data. 

As one can see, differences between the chain Jevons and the chain CSWD indices are negligible, while values of~the chain Dutot price index are considerably more divergent (see Fig. \ref{fig-corrs}). The compared chain weighted indices seem to generate very similar values (see Fig. \ref{fig-chains-two}) with one exception (see Fig. \ref{fig-chains-two} -- coffee on the right). Namely, it seems that high and rapid dynamics of~available products (see Fig. \ref{fig-prod-fraction} -- coffee) can lead to differences even between superlative price indices and their chain versions, although under normal and stable market conditions, these indices approximate each other \citep{diewert1976exact}. In our study, superlative chain indices calculated for coffee differ from each other substantially, e.g. over 2 p.p. for July 2019.

Results presented in Figure \ref{fig-comp-weight} show that the Jevons index tends to generate values a~few percentage points lower than the chain Fisher and the GEKS indexes. The differences between the chain Fisher and the GEKS index values are much smaller, rarely exceeding 1 percentage point, but bigger differences do occur (see Fig. \ref{fig-chains-two}).

\subsection{Differences between full-window multilateral methods}

It is commonly known that values of~multilateral indices do not usually differ strongly; nevertheless, substantial differences between these indices can be observed for very dynamic scanner data sets, including many new and disappearing products \citep{chessa2017comparison}. Figure \ref{fig-chains-two} shows a~comparison of~four popular multilateral price indices calculated for the previously considered data sets.

\begin{figure}[ht!]
    \centering
    \includegraphics[width=0.9\textwidth]{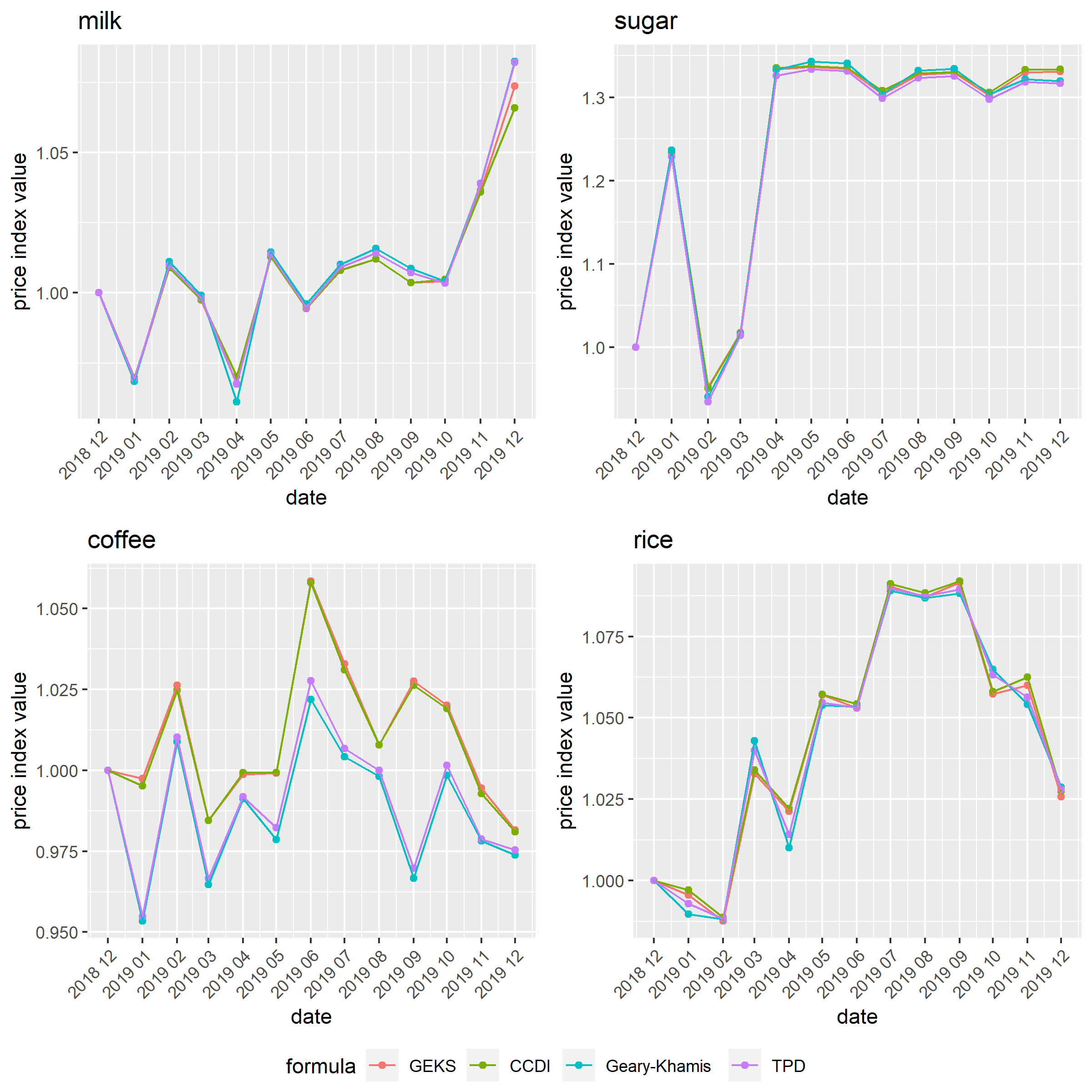}
    \caption{Comparison of~selected full-window multilateral price index indices}
    \label{fig-select-full}
\end{figure}

\begin{figure}[ht!]
    \centering
    \includegraphics[width=0.9\textwidth]{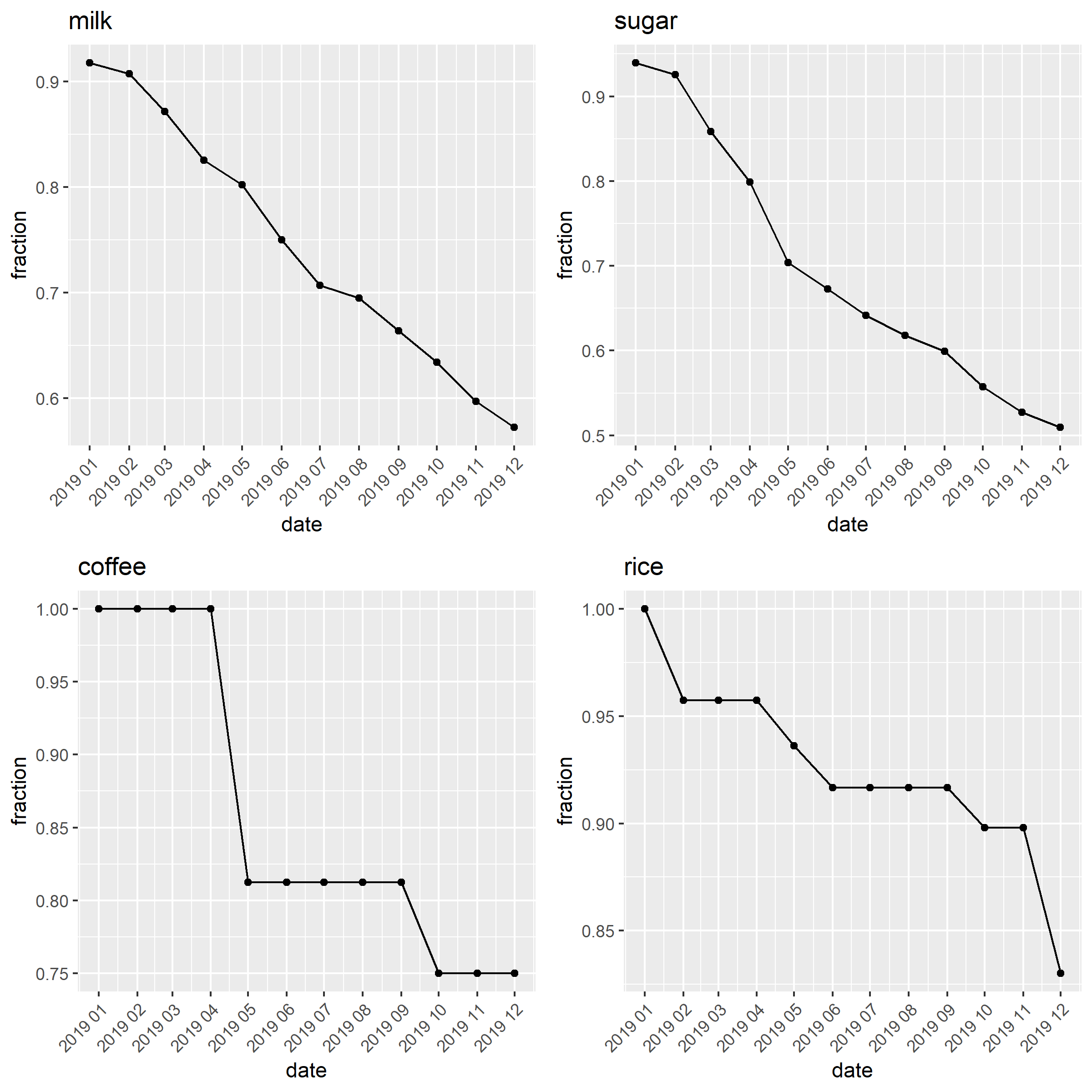}
    \caption{The product fraction remaining on sale since Dec 2018}
    \label{fig-prod-fraction}
\end{figure}

Most results presented in Figure \ref{fig-select-full} confirm the general observation about the similarity of~multilateral indices. However, in the case of~coffee (see Fig. \ref{fig-select-full} -- coffee), while there is hardly any difference between the TPD and the Geary-Khamis indices, the GEKS and CCDI indices deviate somewhat from the values obtained for the other two. This result is similar to that obtained by \citet{chessa2017comparison}, who explain that the main reason for this behavior of~indices is the level of~inflow and outflow rates associated with disappearing and new products. Our results partly confirm this conclusion, but it seems that the speed of~product range changes in the retail chain plays an equally important role. For instance, note that only 52\% of~sugar products survived during the analyzed year (Fig. \ref{fig-prod-fraction} -- sugar), although there are no substantial differences between multilateral indices for this product group (Fig.  \ref{fig-select-full} -- sugar). On the other hand, only 25\% of~coffee products disappeared from the analysed outlets but changes in the product range were rapid (Fig. \ref{fig-prod-fraction} -- coffee). Perhaps this is the reason why we observe substantial differences between index values in this case (see Fig. \ref{fig-prod-fraction} -- coffee). Note that differences between the GEKS and the CCDI indices are negligible, but it is not surprising because superlative indices tend to approximate each other \citet{diewert1976exact}.

\subsection{Differences between splicing indices}

Two main problems encountered when using splicing methods are connected with the choice of~window length and the right splice moment. We focus on these choices in sections \ref{sec-dif-choice-len} and \ref{sec-dif-choice-split}.

\subsubsection{The choice of~window length}\label{sec-dif-choice-len}

\begin{figure}[ht!]
    \centering
    \includegraphics[width=0.9\textwidth]{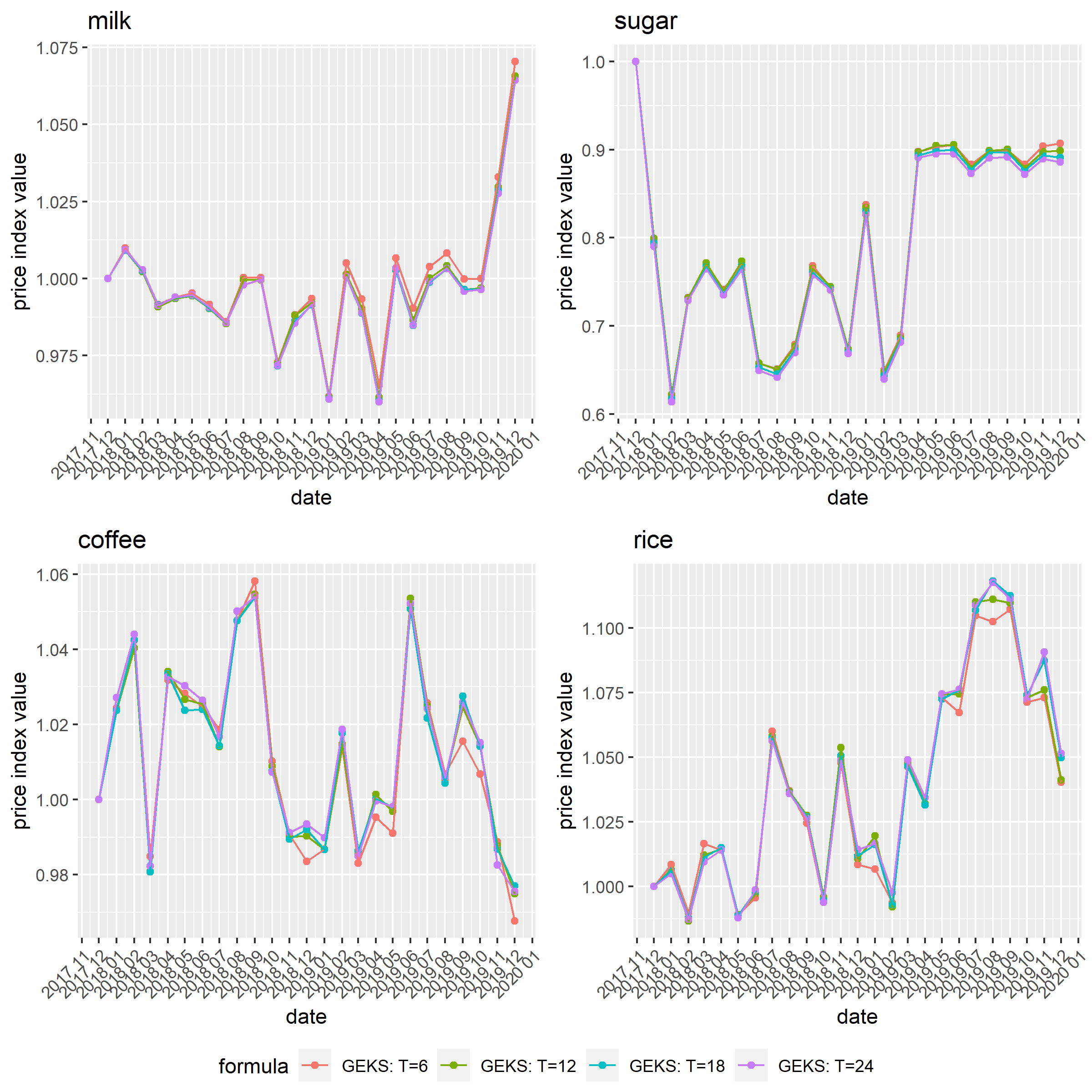}
    \caption{GEKS index values for different values of~window length (the half splice method is used for extensions)}
    \label{fig-geks-index}
\end{figure}

The time window $T+1$  is a~point of~concern. According to Ivanic et.al. (2011), a~13 month window is probably optimal because it is the shortest window that can capture changes in seasonal goods. A longer window would lead to a~loss of~characteristicity \citep{chessa2017comparison}. Although our data sets do not include typical seasonal products, we wanted to investigate the impact of~window length on price index values. The GEKS formula serves as a~benchmark in this analysis and the following values of~window length are considered: 7, 13, 19 and 25 months (each time, December 2017 is the fixed base month). The last case (25 months) covers the complete reference period, i.e. from December 2017 to December 2019. A comparison of~GEKS values for the above-mentioned variants are presented in Figure \ref{fig-geks-index}, showing results for the half splice method (17). Differences between GEKS indices are negligible.

\subsubsection{The choice of~the splicing method}\label{sec-dif-choice-split}

\begin{figure}[ht!]
    \centering
    \includegraphics[width=0.45\textwidth]{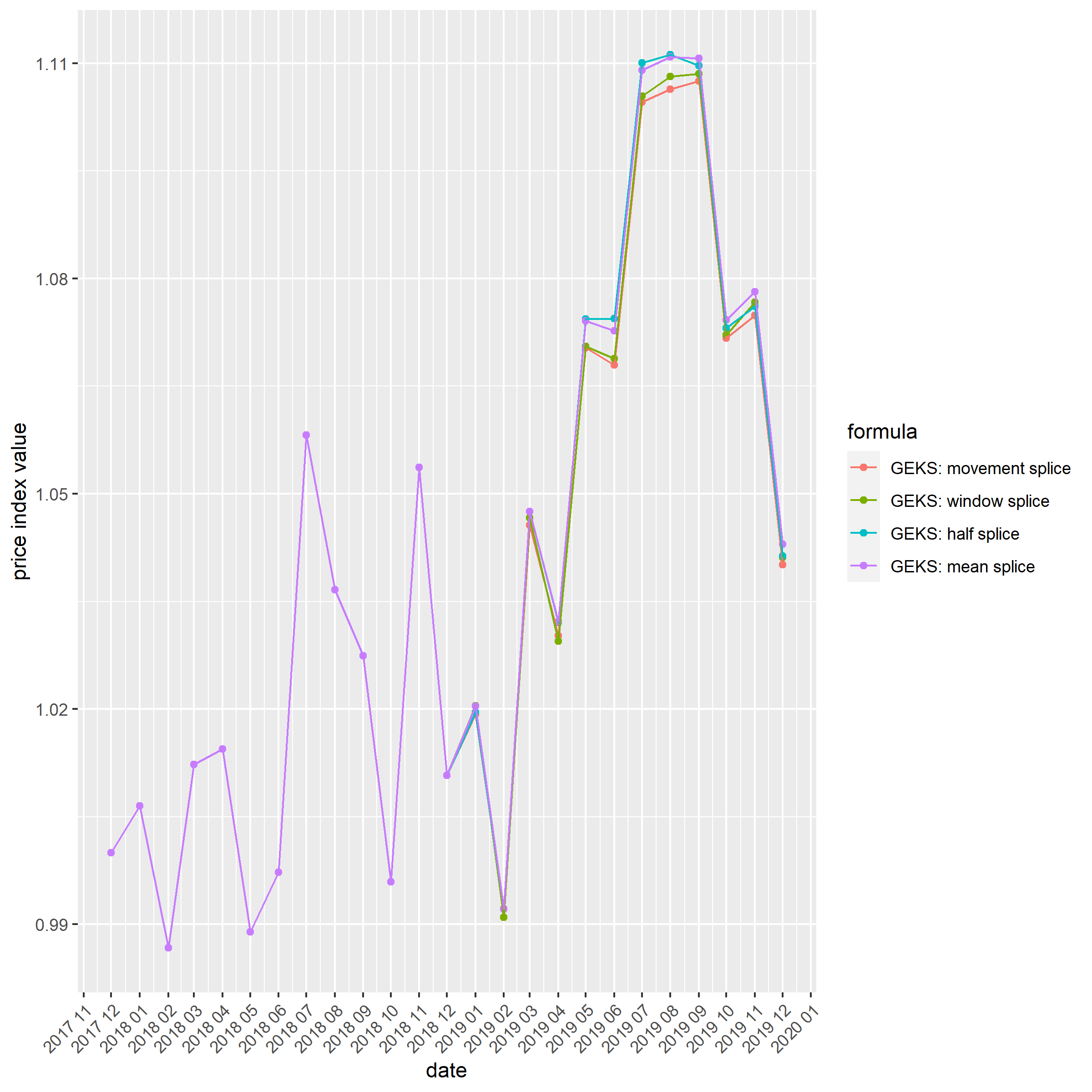}
    \includegraphics[width=0.45\textwidth]{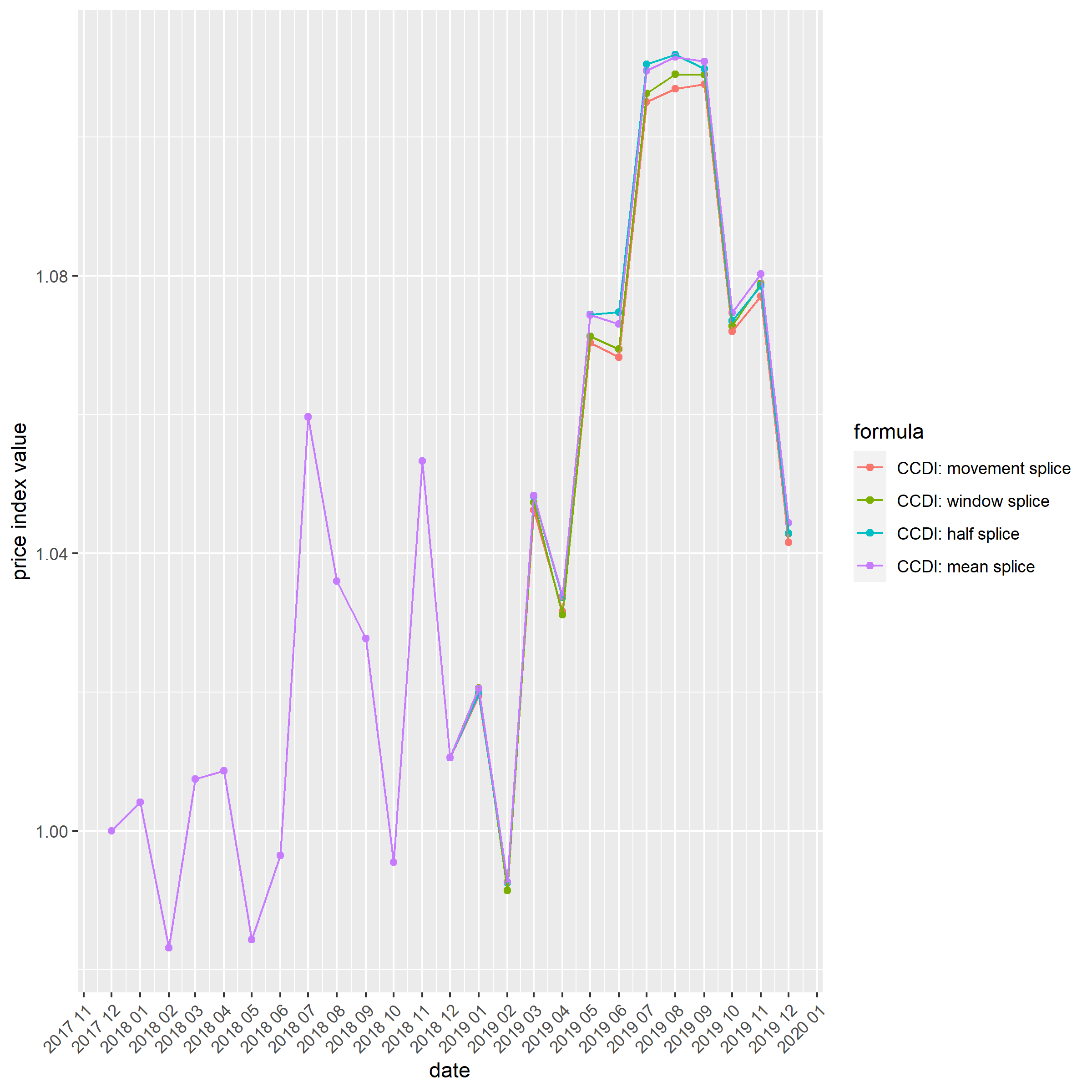}
    \includegraphics[width=0.45\textwidth]{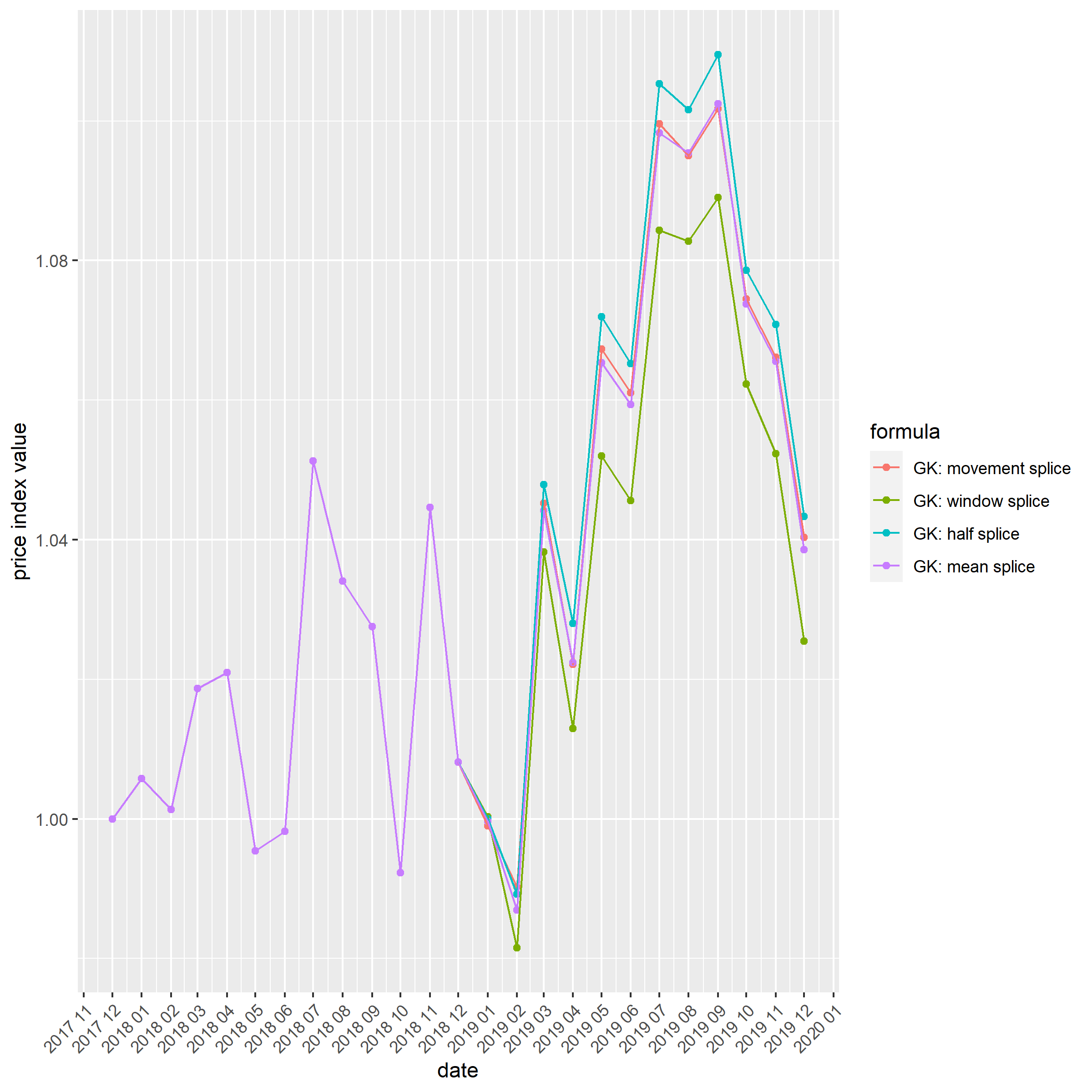}
    \includegraphics[width=0.45\textwidth]{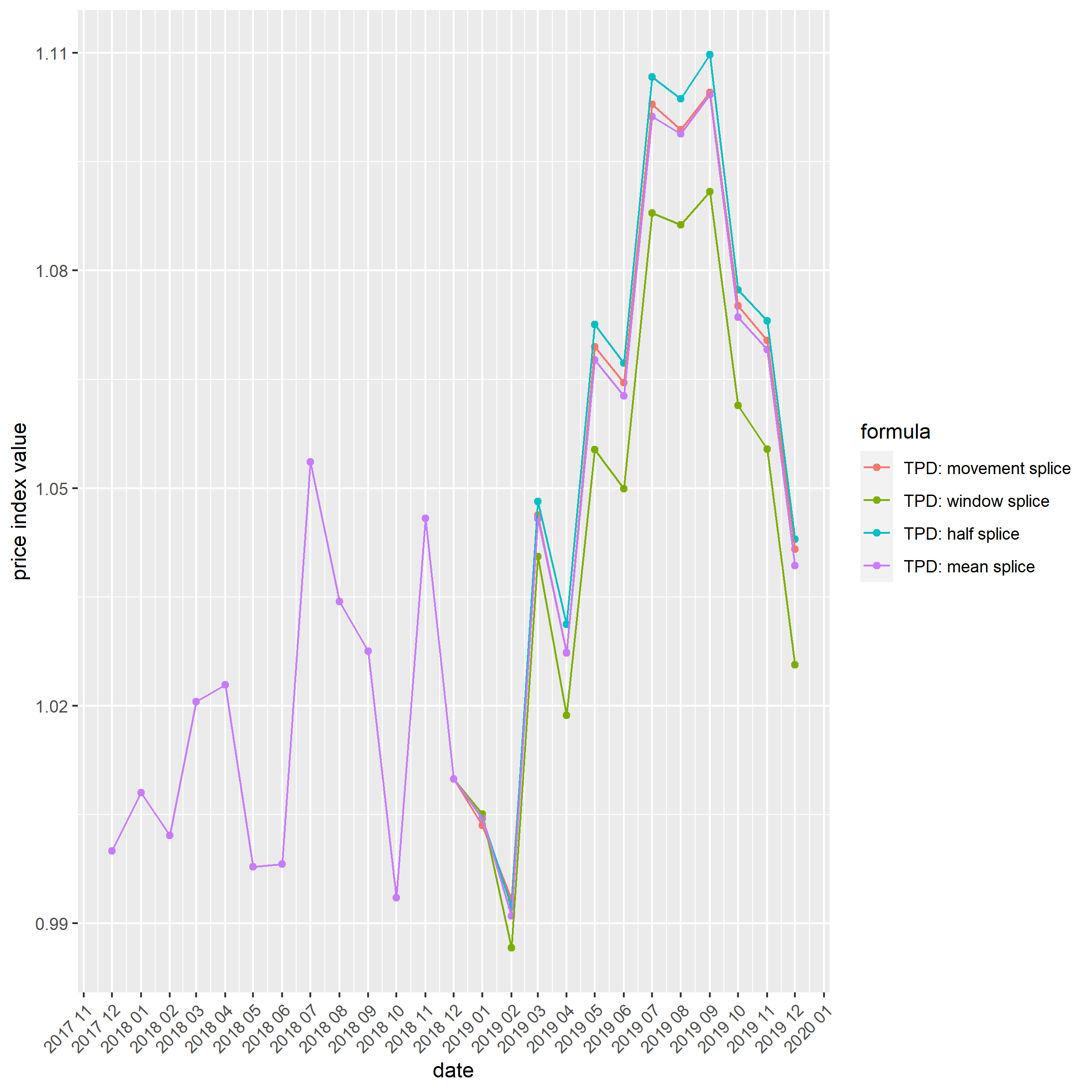}
    \caption{A comparison of~splicing methods for the rice group observed between Dec 2017 and Dec 2019 using a~13-month time window}
    \label{fig-splicing-methods}
\end{figure}

In the following study we analyse the impact of~ the splicing method on values of~multilateral price indices. We consider four indices (GEKS, Geary-Khamis, CCDI and TPD), which are extended by applying all of~the splicing methods discussed in section \ref{sec-splicing-methods}), with a~typical 13-month time window. The index values are calculated for the rice group and compared over the interval from December 2017 to December 2019. The results are presented in Figure \ref{fig-splicing-methods}. 

As can be seen, the Geary-Khamis and the TPD indices seem to be more sensitive to the choice of~the splicing method than the other methods (see Fig. \ref{fig-splicing-methods} bottom row), as evidenced by differences exceeding 1.5 p.p., compared to differences of~less than 0.5 p.p between the GEKS and CCDI indices. Interestingly, on the whole, the window splice and the movement splice methods yield smaller price index values than the other splicing methods. This is consistent with results obtained by \citet{bialek2019}.

\subsection{Extending results by using FBEW and FBMW methods}

\begin{figure}[ht!]
    \centering
    \includegraphics[width=0.7\textwidth]{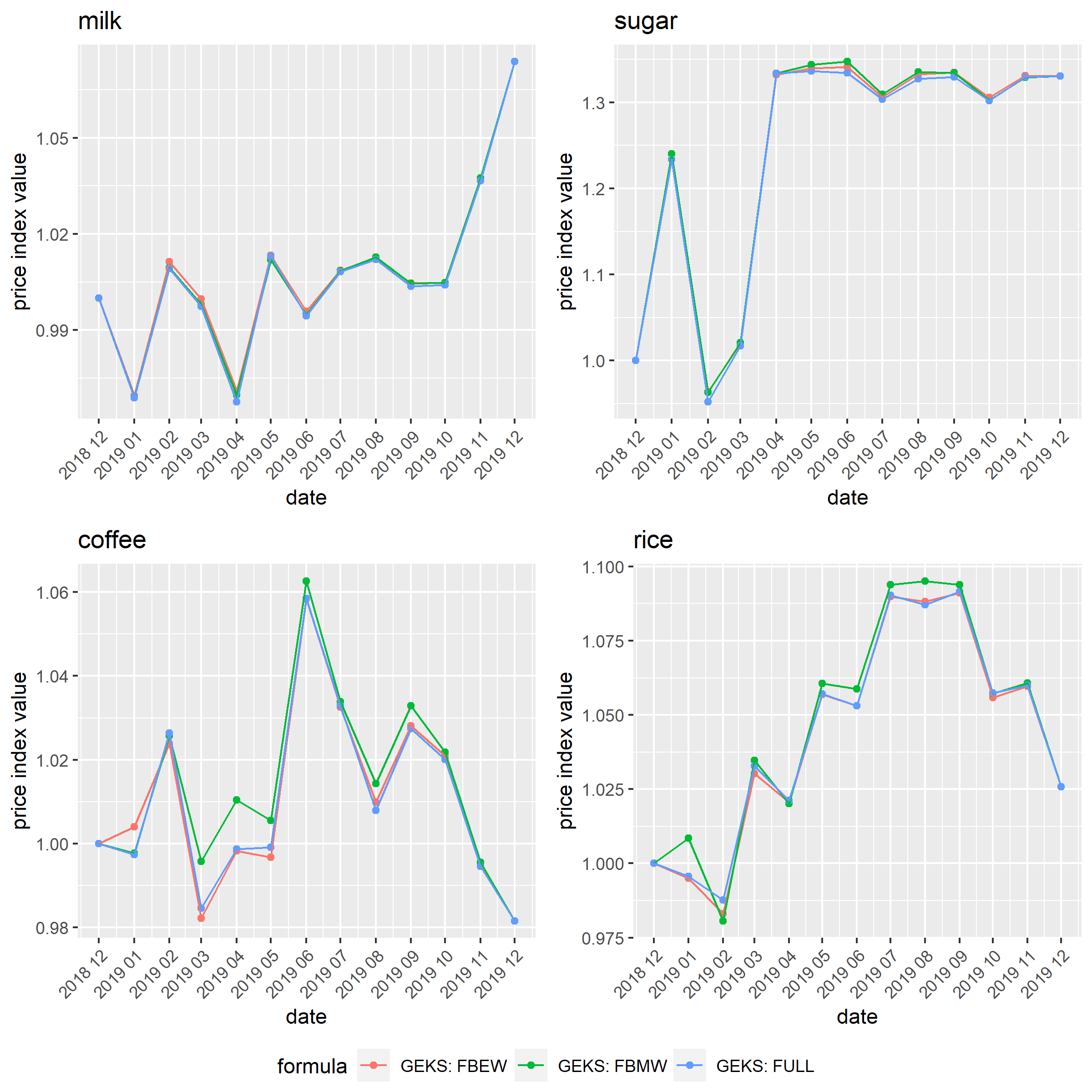}
    \caption{Comparison of~the FBEW, FBMW methods together with the full-window GEKS index}
    \label{fig-comp-fbew-geks}
\end{figure}

There are also other methods for extending multilateral price indices (see section \ref{sec-splicing-methods}). In this part we examine differences between the FBEW and FBMW methods applied to the full-window GEKS index with a~13-month time window (GEKS FULL). The full-window GEKS index serves as a~chain drift free benchmark. The indices are calculated for a~yearly time interval (Dec 2018 – Dec 2019) but in the FBMW method the first time window starts in Dec 2017. December 2018 is the fixed base month. Our results for milk, sugar, coffee  and rice are presented in Figure \ref{fig-comp-fbew-geks}.

As can be expected, the results of~the FBMW method differ from those obtained using the FBEW and GEKS FULL indices since it accounts for prices and quantities from the previous year. In some cases, differences between these methods are negligible (see Fig. \ref{fig-comp-fbew-geks} -- milk and sugar) but sometimes they exceed 1 p.p. (see Fig. \ref{fig-comp-fbew-geks} -- coffee and rice).

\subsection{The choice of~the aggregating formula}

Given the price dynamics for the subgroups that make up a~given COICOP 5 group, calculated for each outlet separately, we are faced with the choice of~how to aggregate this information into a~single price index. In a~traditional data collection, the resultant index for an elementary group is usually calculated as the geometric mean of~indices determined for the component representatives. In the case of~scanner data, given additional information about the level of~consumption, aggregation can also be done using weighted formulas \citep{guerreiro2018use}. In the following part of~the study, three variants of~aggregation are considered: a) aggregation over product subgroups; b) aggregation over outlets; c) double aggregation, i.e. aggregation over product subgroups and over outlets (see sections: \ref{aggr-products}--\ref{aggr-double}). We apply the Laspeyres and the Fisher aggregation formulas and compare them with the option without any aggregation. Note that, in practice, there are at least two higher levels of~aggregation: aggregation over retail chains and final aggregation over data sources, where results obtained from traditional collection and from scanning and web scraping are combined. Results of~aggregation methods for milk, sugar, coffee and rice and for a~yearly time interval are presented in the next three subsections. The GEKS index is treated as the main price index formula, i.e. it is a~measure of~price changes within product subgroups in each outlet separately.

\subsubsection{Aggregation over product subgroups}\label{aggr-products}

\begin{figure}[ht!]
    \centering
    \includegraphics[width=0.7\textwidth]{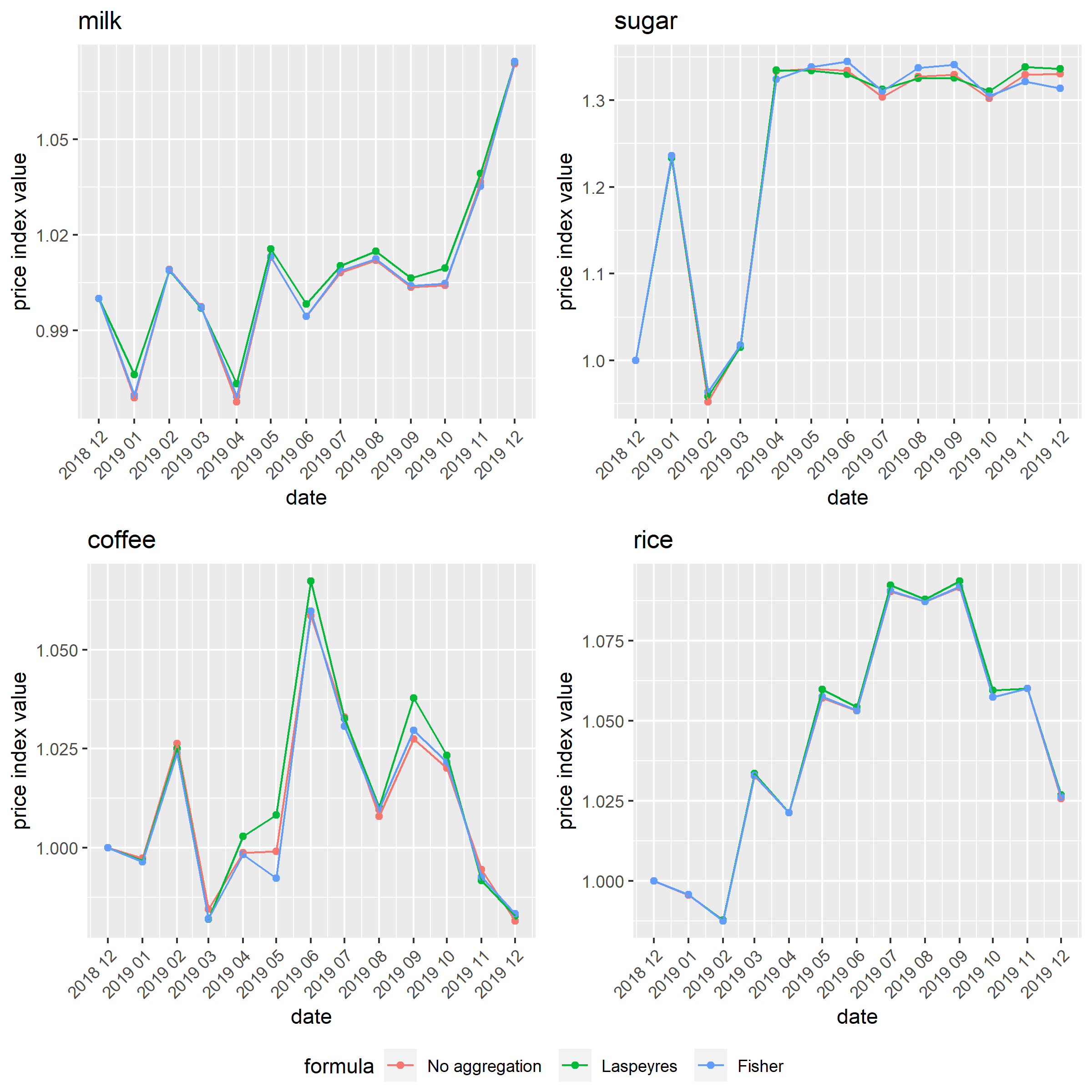}
    \caption{A comparison of~aggregation formulas for the base GEKS index (aggregation over subgroups}
    \label{fig-comp-aggreg-formulas}
\end{figure}

The analysed product groups are COICOP 5 level categories, so they can be further divided into more homogeneous subgroups as follows: milk (fresh whole milk, fresh low-fat milk, concentrated and powdered milk, whole UHT milk, low-fat UHT milk, goat milk), sugar (white sugar, cane sugar, powdered sugar), coffee (instant coffee, coffee beans, ground coffee), rice (long grain rice, white rice). Results of~two formulas involving aggregation over product subgroups are compared in Figure \ref{fig-comp-aggreg-formulas}.

The differences between the two methods when aggregation is done over subgroups are negligible (see Fig. \ref{fig-comp-aggreg-formulas} -- milk, sugar and rice, where differences between index values are less than 0.4 p.p.). However, in the case of~coffee, which includes many new and disappearing products that are replaced rapidly (see Fig. \ref{fig-comp-weight} -- coffee), the choice of~the aggregation method does matter (see Fig. \ref{fig-comp-aggreg-formulas} -- coffee, some differences between index values exceed 1.2 p.p.). 

\subsubsection{Aggregation over outlets}\label{aggr-outlets}

\begin{figure}[ht!]
    \centering
    \includegraphics[width=0.7\textwidth]{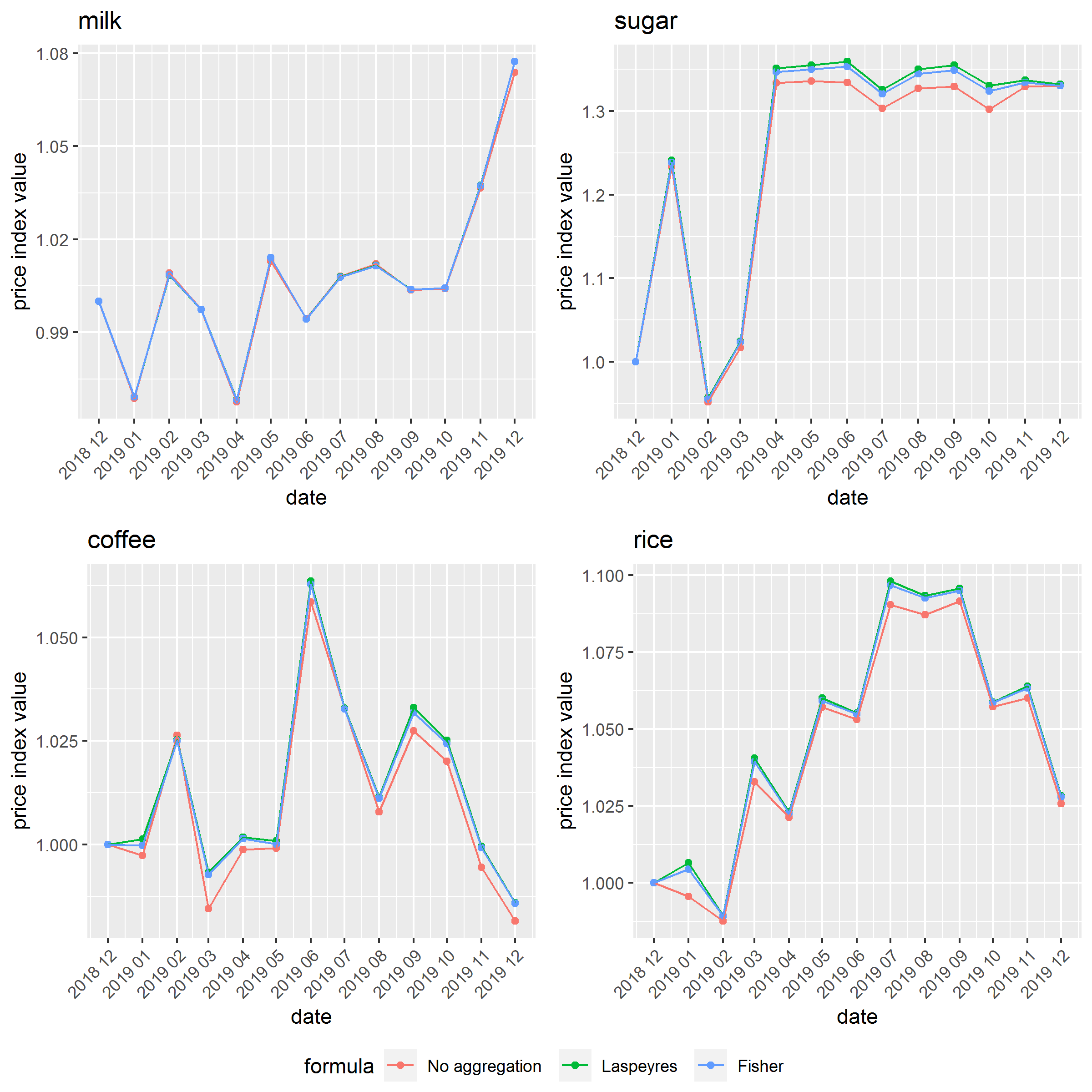}
    \caption{A comparison of~aggregation formulas for the base GEKS index (aggregation over outlets}
    \label{fig-comp-aggreg-formulas-outlets}
\end{figure}

The analysed retail chain sells products in 212 outlets. Results of~two formulas involving aggregation over outlets are compared in Figure \ref{fig-comp-aggreg-formulas-outlets}.

The differences between the aggregation methods (and the option without aggregation) when indices are aggregated over outlets are only negligible for milk (Fig. \ref{fig-comp-aggreg-formulas-outlets}). In the case of~the other product groups, the choice of~the aggregation method makes a~difference - in some cases over 2 percentage points (e.g. Fig. \ref{fig-comp-aggreg-formulas-outlets} sugar) though, for the most part, not bigger than 0.5 p.p.

\subsubsection{Double aggregation}\label{aggr-double}

\begin{figure}[ht!]
    \centering
    \includegraphics[width=0.7\textwidth]{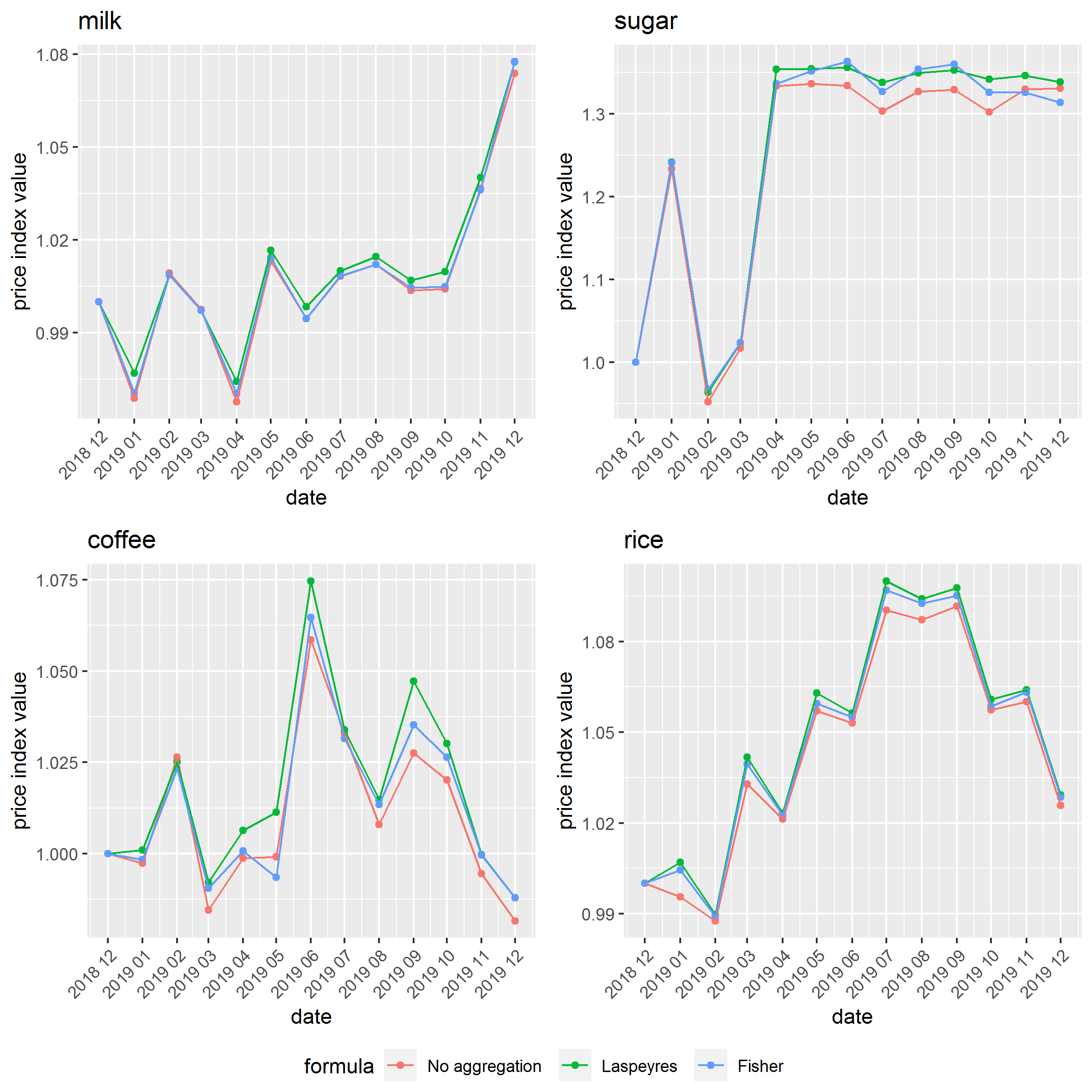}
    \caption{A comparison of~aggregation formulas for the base GEKS index (double aggregation)}
    \label{fig-comp-geks-aggr-a-b}
\end{figure}

Results of~two formulas involving double aggregation, i.e. aggregation over product subgroups and over outlets, are compared in Figure \ref{fig-comp-geks-aggr-a-b}.

When double aggregation is used, the results can be substantially different from those obtained when no aggregation is applied. Also the choice of~the aggregation method matters, as evidenced by the difference between aggregated index values for coffee, which exceeds 1.2 p.p. in September 2019. For the same product group and the same month, the difference between the index value aggregated by means the Laspeyres formula and the non-aggregated index value exceeds 2.4 p.p. (see Fig. \ref{fig-comp-geks-aggr-a-b} coffee). Analogical differences for sugar are even bigger (see Fig. \ref{fig-comp-geks-aggr-a-b} sugar).

\section{Conclusions}\label{sec-conclus}

The potential of~scanner data in the measurement of~inflation is enormous and indisputable, but there are still many unresolved challenges and related problems. As pointed out in the literature, and as we have demonstrated in our study, one of~the main challenges consists in developing appropriate statistical and IT tools for classifying and matching products correctly and automatically. Data preparation, even before any price indices are calculated, is a~complex process, which requires specialized knowledge in the field of~machine-learning and multidimensional statistics. Another important problem is  the choice of~the appropriate price index formula. In the case of~multilateral indices, one must choose window length  and a~method of~extending the index value beyond the selected time window. Finally, a~decision must be made on how to aggregate price indices over individual outlets within a~given retail chain or over potential homogeneous product subgroups. Although Statistics Poland has not yet made its final decision about how to implement scanner data for CPI measurement, conclusions of~our analyses, some of~which are presented in this article, can be regarded as preliminary recommendations for how to solve the above-mentioned problems.

The presented solutions could be considered for regular production of~price indices based on scanner data. Our remarks and preliminary conclusions refer to COICOP 5 product groups described in section \ref{sec-comparison} and can be divided into two groups: suggestions for scanner data processing and remarks concerning the procedure for calculating price indices.  The following remarks and suggestions from the first group can be made: (1) In the case of~analysed scanner data sets (see sections \ref{prepare-data} and \ref{sec-comparison}), text mining involving regular expressions seems to be a reasonable choice for data preparation before the phase of~product classification. (2) Although we have tested many machine-learning methods for classifying products, we have found the Synthetic Minority Over-sampling Technique algorithm and the LASSO algorithm to be the most useful (see Section \ref{product-class-algo}). In particular, prediction based on the LASSO algorithm proved to be sufficiently resistant to new product cases, but it should be emphasised that the high efficiency of~the classification procedure proposed in the article may only be due to the specific nature of~the data sets we worked with. Another thing to bear in mind is that the model parameters, although they retain a~certain stability, need to be re-estimated from time to time, especially when new products appear that have not been included in the data set before. A less effective method but requiring less intervention could involve classification by keywords and phrases defined earlier on the basis of~product labels. (3) The article also presents preliminary results of a two-step procedure for matching products over time (section \ref{sec-matching}). Our approach makes use of EAN codes, product IDs obtained from the retail chain and some product characteristics (if available), such as percentage value, weight, measurement unit and additional product labels. In our opinion, the Jaro-Winkler similarity measure provides best results for matching with blocking (which is available in the \texttt{reclin} R package). It should be noted that the selected value of the scaling factor (0.1) may have been optimal only for our data set. (4) The last stage of~the procedure, just before the calculation of indices, is data filtering. We consider the extreme price filter in two variants (with specific thresholds and with quantile thresholds) and the low sale filter, with different parameter values (section \ref{sec-filtering}). A~general conclusion which can be drawn from our analysis is that unweighted formulas seem to be the most sensitive to the type of data filter and its thresholds. Multilateral indices, by contrast, seem to be less sensitive to the method of data filtering (the article only contains results for the GEKS index but we analysed all the multilateral indices in the study). We also found that the absence of~low sale filtering almost always yields lower values of~price indices, regardless of~the product group. All price index methods are also sensitive to the use of~the quantile version of~the extreme price filter, which means that the level of~quantiles must be set carefully (we used the 1st  and 99th quantile of~all observed price changes to define filter thresholds). 

Our empirical study provides some practical insights about the application of bilateral and multilateral price index methods to scanner data (section \ref{sec-comparison}): (5) Even chain superlative price indices can differ from each other substantially in the case of~dynamic data sets, with a large number of~new and disappearing goods (like the coffee dataset in our study). We write “even” because it is commonly known that superlative indices usually approximate each other. (6) We found that the chain Jevons index, which is commonly applied to scanner data by many NSIs, generates values that are a~few percentage points lower than the chain Fisher and GEKS indices. On the one hand, the Jevons formula is much less problematic from the technical point of~view (see section \ref{sec-weighted}) and is compatible with the traditional measurement of~price dynamics at elementary level. On the other hand, this formula does not account for additional information about expenditures, which can be obtained from scanner data sets and, therefore, seems to be a less adequate choice. In any case, one should realise that replacing unweighted chain formulas with multilateral ones may affect final results considerably. (7) As a~rule, multilateral indices yield similar values and indicate similar price changes. Nevertheless, there are some exceptions. We found that in the case of~dynamic data sets, while there is close similarity between the GEKS index and the CCDI index, and between the Geary-Khamis index and the TPD formula, there are cases where there is discrepancy between these pairs of indices. This behahiour is consistent with some previous results (e.g. \citet{chessa2017comparison}) and has been explained by the level of~inflow and outflow rates associated with new and disappearing goods. However, in our opinion differences between multilateral indices are also strongly affected by the speed of~product range changes. (8) As regards splicing indices, we found that the choice of~window length is not as important as the choice of~the splicing method (sections \ref{sec-dif-choice-len} and \ref{sec-dif-choice-split}). In particular, the Geary-Khamis and the TPD indices seem to be more sensitive to the choice of~the splicing method than the other multilateral indices. In our study, on the whole, the window splice and the movement splice methods generate smaller price index values than the half splice and mean splice methods. (9) With respect to other methods for extending multilateral indices, we found that the FBMW and FBEW methods tend to generate different values with more dynamic data sets. (10) Finally, with regard to the Laspeyres and Fisher formulas for aggregating sub-indices over product subgroups, we noticed that the choice of these aggregation methods has little impact on results in the case of~subcategories for milk, sugar and rice. Nevertheless, aggregation over outlets makes a difference (as does double aggregation), i.e. if data are not aggregated over outlets, the index value can change by a~few tenths of~a~percentage point or even more. For instance, in the case of~sugar (Fig. \ref{fig-comp-aggreg-formulas-outlets}), the GEKS indices calculated for a~variant with and without this type of~aggregation differ from each other by over 2 p.p. 

In summary, the choice of an index formula for scanner data is a~real challenge. We have no doubt that multilateral indices are the optimal choice because they eliminate the effect of~chain drift and additionally take into account the dynamic nature of~scanner data. The Geary-Khamis formula makes it possible to include new products as soon as they appear, which distinguishes this index from other multilateral indices. As far as the best way of extending multilateral indices by splicing, we found the mean splice method to provide the most reliable results without underestimating the real index value. When choosing between the other two methods of~index expansion, i.e. FBEW and FBMW, the latter method seems to be more justified because it always involves a~full (13 month) time window. Taking into account the full time window may, however, be relevant only for products displaying clear seasonality.

\bibliography{main.bib}
\bibliographystyle{apalike}

\end{document}